\documentclass[
aps,prapplied,showpacs,superscriptaddress,numerical,amsmath,amssymb,floatfix,reprint,longbibliography]{revtex4-2}
\usepackage[T1]{fontenc}
\usepackage[utf8]{inputenc}
\usepackage[
pdffitwindow=true,
colorlinks=true,
frenchlinks=false,
linkcolor=blue,
anchorcolor=blue,
citecolor=blue,
filecolor=blue,
urlcolor=blue,
bookmarks=true,
bookmarksopen=true,
bookmarksnumbered=true,
bookmarksopenlevel=1,
plainpages=false,
pdfpagelayout=TwoPageLeft,
pdfpagelabels=true,
breaklinks]{hyperref}
\usepackage[main=english]{babel}
\usepackage[dvipsnames]{xcolor}
\usepackage[per-mode=symbol,separate-uncertainty]{siunitx}
\usepackage{graphicx}
\usepackage{dcolumn}
\usepackage{color}
\usepackage{cleveref}
\usepackage{graphicx,wrapfig,lipsum}
\usepackage{bm}
\usepackage{physics}
\usepackage{color}
\usepackage{ulem}
\usepackage{float}
\usepackage{chemformula}
\usepackage{todonotes}

\Crefname{figure}{Fig.}{Figs.}
\Crefname{equation}{Eq.}{Eqs.}

\begin{document}
\title{Broadband electron paramagnetic resonance spectroscopy of $^{167}$Er:$^{7}$LiYF$_4$ at mK temperatures}
\author{Ana\,Strini\'{c}}
\email[]{Ana.Strinic@wmi.badw.de}
\affiliation{Walther-Mei{\ss}ner-Institut, Bayerische Akademie der Wissenschaften, 85748 Garching, Germany} 
\affiliation{School of Natural Sciences, Technical University of Munich, 85748 Garching, Germany}
\affiliation{Munich Center for Quantum Science and Technology (MCQST), 80799 Munich, Germany}

\author{Patricia\,Oehrl}
\affiliation{Walther-Mei{\ss}ner-Institut, Bayerische Akademie der Wissenschaften, 85748 Garching, Germany} 
\affiliation{School of Natural Sciences, Technical University of Munich, 85748 Garching, Germany}
\affiliation{Munich Center for Quantum Science and Technology (MCQST), 80799 Munich, Germany}

\author{Achim\,Marx}
\affiliation{Walther-Mei{\ss}ner-Institut, Bayerische Akademie der Wissenschaften, 85748 Garching, Germany} 

\author{Pavel\,A.\,Bushev}
\affiliation{Institute of Functional Quantum Systems (PGI-13), Forschungszentrum Jülich, 52425 Jülich, Germany}

\author{Hans\,Huebl}
\affiliation{Walther-Mei{\ss}ner-Institut, Bayerische Akademie der Wissenschaften, 85748 Garching, Germany}
\affiliation{School of Natural Sciences, Technical University of Munich, 85748 Garching, Germany}
\affiliation{Munich Center for Quantum Science and Technology (MCQST), 80799 Munich, Germany}

\author{Rudolf\,Gross}
\affiliation{Walther-Mei{\ss}ner-Institut, Bayerische Akademie der Wissenschaften, 85748 Garching, Germany}
\affiliation{School of Natural Sciences, Technical University of Munich, 85748 Garching, Germany}
\affiliation{Munich Center for Quantum Science and Technology (MCQST), 80799 Munich, Germany}

\author{Nadezhda\,Kukharchyk}
\email[]{Nadezhda.Kukharchyk@wmi.badw.de}
\affiliation{Walther-Mei{\ss}ner-Institut, Bayerische Akademie der Wissenschaften, 85748 Garching, Germany}
\affiliation{School of Natural Sciences, Technical University of Munich, 85748 Garching, Germany}
\affiliation{Munich Center for Quantum Science and Technology (MCQST), 80799 Munich, Germany}

\date{\today}
\pacs{}
\keywords{}  

\begin{abstract}

Rare-earth spin ensembles are a promising platform for microwave quantum memory applications due to their hyperfine transitions, which can exhibit exceptionally long coherence times when using an operation point with zero first-order Zeeman (ZEFOZ) shift. 
In this work, we use broadband electron paramagnetic resonance (EPR) spectroscopy on $^{167}$Er:$^{7}$LiYF$_4$ single crystals at sub-Kelvin temperatures. By fitting the spin Hamiltonian to the zero-field spectrum, we obtain refined parameters of the magnetic field-independent interactions, such as the hyperfine and quadrupole interaction. We also study the influence of the quadrupole interaction on the hyperfine splitting in the zero and low magnetic field range by analyzing EPR-spectra between \SI{0}{\milli\tesla} and \SI{50}{\milli\tesla}.
Our findings highlight the broadband EPR spectroscopy approach as a powerful tool for the precise determination of the spin Hamiltonian parameters and for the characterization of hyperfine transitions in terms of their selection rules and linewidth.

\end{abstract}

\maketitle
\section{Introduction} 
Hybrid quantum systems consisting of a quantum processor coupled to a quantum memory possess great potential for quantum computing~\cite{Kurizki2015, Gouzien2021}. To this end, efficient interfacing of the quantum processor to the quantum memory is essential, implying direct compatibility of their operation conditions and enabling a direct low-loss coupling between both devices. Superconducting quantum processors are operated at GHz-frequencies, millikelvin temperatures, and a low magnetic field environment~\cite{Krantz2019, Schneider2019}. These operation conditions are compatible with several spin-ensemble platforms, such as nitrogen-vacancy color centers in diamond~\cite{Kubo2011, Grezes2014}, donors in silicon~\cite{OSullivan2020, Zollitsch2015, Weichselbaumer2020}, and rare-earth doped crystals~\cite{Wolfowicz2015, Afzelius2013, Probst2015}. Therefore, these systems are promising candidates for establishing microwave quantum memory devices. 

Rare-earth spin ensembles are particularly promising for interfacing microwave quantum memories with superconducting quantum processors based on qubits operating at microwave frequencies. First, they are frequency-matched due to their zero-field hyperfine transitions in the GHz-frequency range, and second, they can exhibit exceptionally long spin coherence times up to a few milliseconds~\cite{LeDantec2021}. These long coherence times are possible by making use of dynamical decoupling pulse sequences~\cite{Merkel2021}, probing ultra-pure crystals \cite{LeDantec2021}, or utilizing transitions with so-called zero first-order Zeeman (ZEFOZ) shift. In the literature, the latter are also referred to as 'clock transitions'~\cite{Wolfowicz2013, Ortu2018}. As ZEFOZ transitions are, to first order, insensitive to magnetic field fluctuations, significantly increased coherence times can be achieved due to the reduced impact of magnetic field noise~\cite{ McAuslan2012}. They are inherent to many hyperfine transitions of rare-earth ions embedded in various host materials and can be found even at zero and low magnetic field strength~\cite{Rakonjac2020, Ortu2018}. However, finding true ZEFOZ transitions can be experimentally challenging due to the complex hyperfine level structure of rare-earth ions~\cite{Alizadeh2023}.
In the same way, the exact theoretical modelling of the hyperfine level structure, allowing to identify transitions suitable for microwave quantum memory applications, requires the precise experimental determination of the parameters as well as the identification of all terms entering the spin Hamiltonian. 

Electron paramagnetic resonance (EPR) spectroscopy is a powerful experimental approach to characterize the magnetic properties of electronic spins in solids with high precision. Typically, EPR studies are performed by coupling the spin system to a microwave cavity or resonator and measuring the absorption of a microwave signal while sweeping the externally applied magnetic field. At resonance, the coupling of the spin systems to the resonant mode of the cavity can be detected as a change in the real and imaginary parts of the microwave signal. However, this field-swept approach is limited to a narrow frequency range determined by the narrow bandwidth of the microwave resonator. Therefore, in recent years, different broadband techniques have been developed, allowing for sweeping both the magnetic field and microwave frequency and thereby accessing a broad parameter regime. A particularly powerful technique is broadband EPR spectroscopy based on a coplanar transmission line~\cite{Clauss2013, Prinz-Zwick2022, Wiemann2015, Jing2019, Miksch2020}. Here, the spin system is placed on top of the transmission line, and the microwave transmission and/or reflection is measured as a function of the applied magnetic field and microwave frequency over a wide range using a vector network analyzer. Such a technique is common for broadband ferromagnetic resonance experiments \cite{MaierFlaig2017, Klingler2017}. 

In this work, we perform broadband EPR spectroscopy on a diluted $^{167}$Er:$^{7}$LiYF$_4$ crystal at millikelvin temperatures to identify the hyperfine transitions of $^{167}$Er. We precisely determine the hyperfine and quadrupole parameters of the spin Hamiltonian using a least squares fit of the measured zero-field spectrum. Moreover, based on the broadband spectra recorded in the magnetic field range from \SI{0}{\milli\tesla} up to \SI{50}{\milli\tesla}, with the magnetic field $\boldsymbol{B_{0}}$ aligned along different crystallographic directions of the LiYF$_{4}$ host crystal, we can confirm the already published $g$-factors. By taking spectra in Faraday- and Voigt-like geometry, we can also probe different selection rules, which we can reliably simulate. We further determine the effective temperature of the spin ensemble. Finally, the broadband spectra allow us to extract the linewidth of the hyperfine transitions. We used this data to get insight into the linewidth narrowing of the hyperfine transitions around their ZEFOZ points.

\section{Experimental Techniques}

\begin{figure}[tbh]
\includegraphics[width=0.48\textwidth]{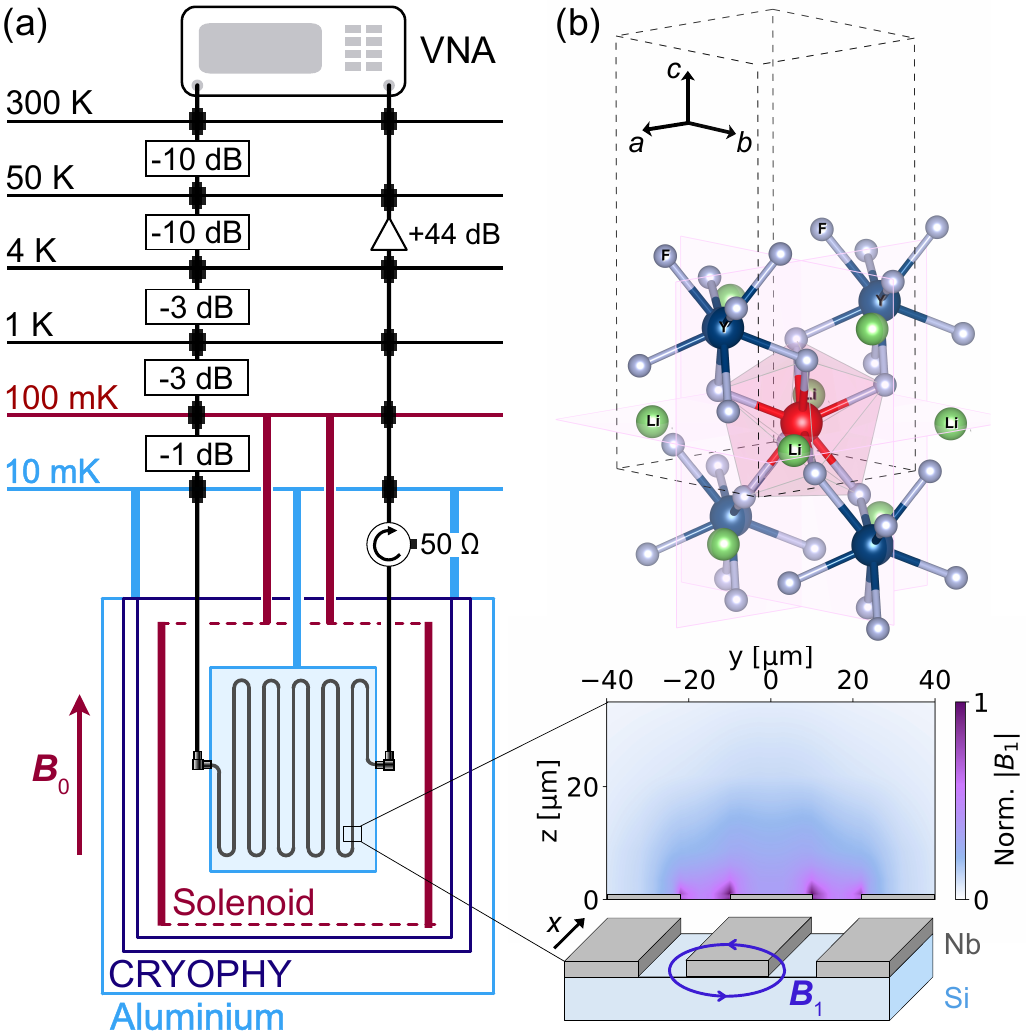}
\caption{(a)~Schematics of the measurement setup indicating the thermal stages of the dilution refrigerator and the microwave wiring. The studied crystal is placed on top of a meander-shaped CPW in the bore of a superconducting coil producing a magnetic field $B_{0}$. A VNA is used to measure the complex transmission parameter S$_{21}$. Zoom-in: simulated distribution of the propagating magnetic mode $B_{1}$ around the center conductor of the CPW with a current flow in $x$ direction, normalized to its maximum value. (b)~Coordination of an Er$^{3+}$-ion (red) within the unit cell of $^{7}$LiYF$_4$ based on published lattice parameters~\cite{Momma2011, Goryunov1992}.}
\label{fig:fig1}
\end{figure}

The experimental setup used for the broadband EPR spectroscopy is schematically shown in \Cref{fig:fig1}~(a). We study a $^{167}$Er:$^{7}$LiYF$_4$ crystal with an $^{167}$Er concentration of 0.0025\%, which was grown using the Bridgeman-Stockager technique (see Ref.~\cite{Kukharchyk2018}). 

The crystal is placed on top of a meander-shaped superconducting coplanar waveguide (CPW). 
The superconducting circuit is fabricated using niobium sputter deposition on a high-resistivity silicon substrate and subsequent patterning by electron beam lithography and reactive ion etching. 
The geometry of the CPW with a centre conductor width of \SI{20}{\micro\meter} and a gap width of \SI{12}{\micro\meter} also allows us to simulate the distribution of the microwave $B_{1}$ excitation field in the proximity of the CPW [see zoom-in in \Cref{fig:fig1}(a)]. The spatial dependence of the normalized magnitude $|B_{1}|$ highlights the rapid decay of the signal strength with increasing distance from the CPW surface.
Therefore, we choose a meander-shaped design to effectively couple the spins to the propagating microwave field and probe a large crystal volume. The CPW is wire-bonded via taper patches to a gold-plated printed circuit board, interfaced to microwave coaxial cables with solderless, non-magnetic SMA connectors. Throughout the microwave assembly, we aim for an impedance matching of \SI{50}{\ohm}. 

The sample assembly is thermally anchored to the mixing-chamber stage of a dry dilution refrigerator (Bluefors, XLD~1000) with a base temperature of less than $\SI{10}{\milli\kelvin}$. It is positioned within the bore of a home-built superconducting (NbTi) solenoid magnet~\cite{Vogl2023}, which is thermally anchored to the \SI{100}{\milli\kelvin} stage of the dilution refrigerator and can generate a static magnetic field ($B_{0}$) up to $\SI{\pm50}{\milli\tesla}$. In addition, the magnet is enclosed by three layers of magnetic shielding. On the one hand, the magnetic shielding confines the magnetic field within the coil, allowing for compatibility with superconducting qubits operated within the same cryostat. On the other hand, the magnitude of the background magnetic fields at the sample position is strongly reduced, which is crucial for ultra-low field EPR spectroscopy. 

To probe the selection rules, the spectroscopy measurements have been performed for $\boldsymbol{B_{0}}$ aligned both parallel and perpendicular to the microwave propagation vector $\boldsymbol{k}$. In the following, we adopt the common nomenclature used for propagating wave experiments and refer to the probed excitation geometries as Faraday-like geometry ($\boldsymbol{k} \parallel \boldsymbol{B_{0}}$) and Voigt-like geometry ($\boldsymbol{k} \perp \boldsymbol{B_{0}}$) \cite{Nehrkorn2015}. Further details of the positioning of the sample on top of the CPW and its alignment are discussed in \Cref{appendix:b}.

The complex microwave transmission $S_{21}$ is measured with a vector network analyzer (VNA). 
To thermalize the microwave signal with input power $P_\text{{in}}=-50$~dBm to mK temperatures, we attenuate the input line in the dilution refrigerator using a set of attenuators (see \Cref{fig:fig1}~(a)) with a total attenuation of \SI{27}{\dB}. 
The output line is configured with a microwave circulator to reduce the impact of thermal microwave photons from stages with higher temperatures. We use a low-noise HEMT amplifier at the \SI{4}{\kelvin} stage for the pre-amplification of the output signal.
To analyze the magnetic field dependence, we calibrate the applied magnetic field strength via the EPR transition of a 2,2-Diphenyl-1-Picrylhydrazyl (DPPH) sample (Sigma Aldrich) as described in \cref{appendix:a}.
\section{Theoretical Background} \label{sec:theory}

The host crystal $^{7}$LiYF$_4$ belongs to the group of Scheelite crystals with a tetragonal unit cell \cite{Sattler1971}.
When doped with Er$^{3+}$-ions, these substitute the Y$^{3+}$-ions while preserving the local S$_4$ symmetry~\cite{Chukalina1999, Goryunov1992}. The coordination of the Er$^{3+}$-ion within the $^{7}$LiYF$_4$ unit cell is indicated in \Cref{fig:fig1}~(b). The doping induces low strain due to the similar ionic radii of Er$^{3+}$ and Y$^{3+}$~\cite{Vishwamittar1974} and does not lead to any charge compensation as both ions are in the trivalent charge state~\cite{Chukalina1999}. Exceptionally narrow inhomogeneous broadening has been observed on optical transitions of Er:$^{7}$LiYF$_4$. This was attributed to the fact that the influence of crystal defects and strain in isotopically purified $^{7}$LiYF$_4$ is so small that the inhomogeneous broadening becomes dominated by the magnetic field fluctuations produced by nuclear spins of fluorine ions~\cite{Macfarlane1992, Kukharchyk2018}. 

The Er$^{3+}$ ([Xe]4$f^{11}$) belongs to the group of Kramers ions. At cryogenic temperatures, only the lowest Kramers doublet ($J = \frac{15}{2}$) of the ground state (spectroscopic notation: $^4I_{15/2}$) is populated, and the spin system can be approximated with an effective spin $S = 1/2$ \cite{AbragamBleaney}.
In this case, the energy level splitting of the $^{167}$Er$^{3+}$ ($I = 7/2$) ground state can be described by a spin Hamiltonian containing three main terms: 
the electron Zeeman interaction ($H_\text{EZ}$), 
the hyperfine interaction ($H_\text{HF}$) 
and, due to the nuclear spin $I > 1/2$, the nuclear quadrupole interaction ($H_\text{Q}$). The corresponding characteristic interaction tensors $\boldsymbol{g}$ and $\boldsymbol{A}$ are compatible with the tetragonal crystal symmetry and, therefore, their components can be expressed as parallel, $g_{\parallel}$ and $A_{\parallel}$, and perpendicular, $g_{\perp}$ and $A_{\perp}$, to the crystallographic $c$-axis, when presented in diagonal form~\cite{AbragamBleaney}. Assuming axial symmetry, the quadrupole interaction tensor $\boldsymbol{Q}$ is traceless and can be described by a single parameter $P$, which relates to the largest principal value of the electric field gradient at the nucleus \cite{Stoll2017}. 
The spin Hamiltonian can be expressed as follows: 
\begin{equation}
\begin{split}
    H_\text{eff} = & H_\text{EZ} + H_\text{HF} + H_\text{Q} \\
     = & \mu_\text{B} g_{\perp} (B_{x} \boldsymbol{S_{x}} + B_{y} \boldsymbol{S_{y}}) + \mu_\text{B} g_{\parallel} B_{z} \boldsymbol{S_{z}}\\
     & + A_{\perp} (\boldsymbol{S_{x}} \boldsymbol{I_{x}} + \boldsymbol{S_{y}} \boldsymbol{I_{y}}) + A_{\parallel} \boldsymbol{S_{z}} \boldsymbol{I_{z}}\\
    & + P (3\boldsymbol{I_{z}}^{2} +  \boldsymbol{I}^{2}),\\
\end{split}
\label{equation:1}
\end{equation}
with the Bohr magneton $\mu_\text{B}$, the applied magnetic field $\boldsymbol{B}=(B_x,B_y,B_z)$, and the electron ($\boldsymbol{S_{x}} , \boldsymbol{S_{y}} , \boldsymbol{S_{z}}$) and nuclear ($\boldsymbol{I_{x}} , \boldsymbol{I_{y}} , \boldsymbol{I_{z}}$) spin operators. By diagonalizing the spin Hamiltonian in~\Cref{equation:1} numerically, the eigenenergies and eigenstates, as well as the related hyperfine transitions between these states, are obtained. The transition probability between two states $\ket{i}$ and $\ket{f}$ under the action of a microwave magnetic field $\boldsymbol{B_{1}}$ is determined by the matrix element of the interaction energy $H_\text{1} = - \mu_\text{B} \boldsymbol{B_{1}} \boldsymbol{g} \hat{\boldsymbol{S}}$ \cite{Stoll2014}. The intensity $I_{if}$ of the corresponding spectroscopic line observed in an EPR experiment is proportional to
\begin{equation}
    I_{if} \propto  \chi \; |\mel{f}{H_{1} } {i}|^2 \; .
\label{equation:2}
\end{equation}
Here, $\chi$ is the population difference between the initial and final state, which is given by a Boltzmann distribution in thermal equilibrium, as described in \Cref{{appendix:c}}. By choosing $\boldsymbol{B_{1}}$ according to different excitation geometries realized in the experiments, the selection rules and transition probabilities can be simulated. Details on the experimental configuration, as well as the analysis of the selection rules and transition probabilities, are given in~\Cref{Sec:Magnetic field}.

Previous studies on Er:LiYF$_4$ utilized different experimental techniques to determine the parameters entering the spin Hamiltonian. They include cavity-based EPR spectroscopy~\cite{Brown1969, Sattler1971, Guedes2002, Marino2016}, polarized excitation spectra~\cite{Macfarlane1992}, magneto-optical spectroscopy~\cite{Lisin2019,Gerasimov2016}, optical vector network analysis~\cite{Kukharchyk2020}, and Fourier-transform spectroscopy~\cite{Popova2000}. The published spin Hamiltonian parameters for Er:LiYF$_4$ are summarized in \Cref{table:1}, excluding Refs.~\cite{Gerasimov2016, Popova2000} as these provide the crystal field Hamiltonian parameters. In references specifying the spin Hamiltonian parameters, the quadrupole interaction has been neglected. As shown below, this leads to a significant discrepancy between theory and experiment, particularly in the low magnetic field range.
In contrast, studies based on the crystal field calculations usually include the quadrupole interaction~\cite{Gerasimov2016, Popova2000}. However, they do not provide the parameters entering the spin Hamiltonian and hence do not allow for any comparison to the parameters obtained in this work. A particular strength of our study is the fact that all parameters determining the splitting in the absence of a magnetic field can be derived in a single experiment with high precision. Obviously, for determining $g_\perp$ and $g_\parallel$, at least two experiments with different orientations of the crystal are needed.

\begin{table}[tbh]
\centering
\begin{tabular}{ |c|c|c|c|c|c| } 
\hline
Source & $g_\parallel$ & $g_\perp$ & $A_\parallel$ (MHz) & $A_\perp$ (MHz) & $P$ (MHz) \\
\hline
\cite{Brown1969} & +~3.32 & 8.09 & - & - & - \\ 
\hline
\cite{Sattler1971} & 3.137 & 8.105 & 325.8 & 840 & - \\ 
\hline
\cite{Macfarlane1992} & 3.132 & 8.102 & - & - & - \\ 
\hline
\cite{Guedes2002} & 3.130 & 7.929 & 325 & 816 & - \\
\hline
\cite{Wu2004} & 3.141 & 7.932 & 334* & 824* & - \\
\hline
\cite{Marino2016} & 3.147 & 8.105 & - & - & - \\ 
\hline
\cite{Gerasimov2016} & 3.132 & 8.102 & - & - & -\\ 
\hline
\cite{Lisin2019}, \cite{Kukharchyk2020} & 3.137 & 8.1 & -325.8 & 840 & -  \\ 
\hline
this work & 3.137 & 8.105 & -319.6 & -844.2 & -7.184 \\ 
\hline
\end{tabular}
\caption{Summary of the published spin Hamiltonian parameters for Er:LiYF$_4$ as well as those determined in this work. 
*Ref.~\cite{Wu2004} specifies $A_\parallel$ and $A_\perp$ in units of $10^4$~cm$^{-1}$, while \SI{}{\mega\hertz} seems more appropriate.}
\label{table:1}
\end{table}

\section{Results and Discussion} 
\subsection{Background subtraction}

As described above, we determine the microwave transmission across our setup by measuring the complex scattering parameter $S_{21}$ as a function of frequency at a constant applied magnetic field via a VNA. Besides information on the sample properties, the measured $S_{21}(f)$ dependence is also affected by the characteristic spectral features of the individual microwave components included in the microwave path, caused, e.g., by a finite impedance mismatch. While such contributions can be removed by a thorough microwave calibration, this is very challenging in practice due to the complexity of the setup and the wide frequency range of our experiments. 
To disentangle the signal contributions due to the spin system from those related to the microwave setup, we follow the simple approach of subtracting a reference background $\abs{S_{21,\text{ref}}(f)}$ from the microwave spectrum $\abs{S_{21,B_{i}}(f)}$ at each magnetic field. Similar approaches for subtraction-based background removal were used in \cite{Prinz-Zwick2022, Wiemann2015}. As our magnetic field range is limited to $B_{0} = \SI{50}{\milli\tesla}$, it is not possible to take a reference spectrum at a magnetic field strength where no spin absorption is present within the frequency range of interest. Therefore, we construct the reference spectrum $S_{21,\text{ref}}(f)$ by selecting specific parts of the spectra ($S_{21,B_{i}}(f)$), which have been measured at different magnetic field strengths where no absorption by the spin system is present. In this way, a reference background $\abs{S_{21,\text{ref}}(f)}$ is constructed, representing the transmission pattern of the overall microwave circuit.
Accordingly, we obtain the background-corrected microwave transmission to 
\begin{equation}
   \Delta\abs{S_{21,B_i}(f)} = \abs{S_{21,\text{ref}}(f)} - \abs{S_{21,B_{i}}(f)} \; .
    \label{equation:4}
\end{equation} 
Note that $\Delta\abs{S_{21}(f)}$ is related only to the effective change in microwave transmission due to the absorption by the spin ensemble, i.e., it contains only the magnetic field-dependent information, which we associate with the spin system. A more detailed discussion on how the background removal technique can affect the extracted parameters is given in \Cref{appendix:d}.

\subsection{Zero magnetic field spectroscopy}

\begin{figure*}
\includegraphics[width=0.99\textwidth]{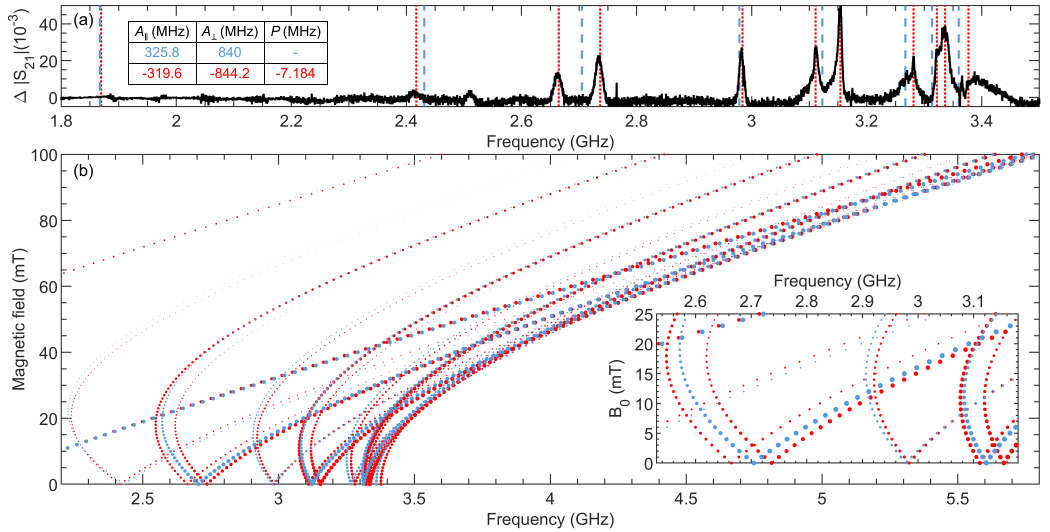}%
\caption{(a)~Background-subtracted zero-field spectrum (black) measured between \SI{1.8}{\giga\hertz} and \SI{3.5}{\giga\hertz}. The blue dashed and the red dotted vertical lines mark the simulated transition frequencies utilizing the published spin Hamiltonian parameters \cite{Sattler1971} and those obtained by fitting our experimental data, respectively. (b)~Simulated magnetic field dependence of the hyperfine transition frequencies for $\boldsymbol{B_{0}} \parallel \boldsymbol{c}$ using the same color code as in (a). The inset shows a zoom-in of the frequency range from \SI{2.5}{\giga\hertz} to \SI{3.2}{\giga\hertz} and magnetic field range from \SI{0}{\milli\tesla} to \SI{25}{\milli\tesla}. Note that neglecting the quadrupole interaction parameter $P$ leads to a significant discrepancy between experimental and calculated transition frequencies particularly in the low-field regime.
}
\label{fig:fig2}
\end{figure*}

In the following, the term 'zero magnetic field spectroscopy' indicates that we perform microwave transmission experiments at zero applied static magnetic field, $B_{0} =\SI{0}{\milli\tesla}$. At the same time, the earth's magnetic field is attenuated by three layers of magnetic shielding down to approx.~\SI{50}{\nano\tesla}~\cite{Vogl2023}. The background-subtracted zero-field spectrum measured in the frequency range from \SI{1.8}{\giga\hertz} to \SI{3.5}{\giga\hertz} is plotted in \Cref{fig:fig2} (a) as the black solid line.
To extract the parameters entering the spin Hamiltonian from this data, we numerically calculate the corresponding frequencies of the allowed EPR transitions as described in Sec.~\ref{sec:theory}.
Since the hyperfine and quadrupole interactions are field-independent, they determine the energy level splitting at zero magnetic field, while the Zeeman interaction dominates at higher magnetic field strengths. 
We extract the refined quadrupole and hyperfine parameters via a non-linear least squares fitting routine of the calculated transition frequencies to the measured zero-field spectrum. We obtain $A_{\parallel} = \SI{-319.6}{\mega\hertz} \pm \SI{8.68}{\kilo\hertz}$, $A_{\perp} = \SI{-844.2}{\mega\hertz} \pm \SI{9.44}{\kilo\hertz}$ and $P = \SI{-7.184}{\mega\hertz} \pm \SI{1.09}{\kilo\hertz}$, with the error of the respective parameter indicating the extracted standard deviation from the fitting routine.
Furthermore, we find the ratio $g_{\parallel} A_{\perp} / g_{\perp} A_{\parallel} = 1.023$, indicating that $J$-state mixing can be neglected~\cite{Sattler1971, Elliott1953, Abraham1983}. The negative sign of $A_{\parallel}$, $A_{\perp}$ and $P$ is chosen for various reasons. First, it matches the symmetry rule $\frac{A_{\parallel}}{g_{\parallel}} = \frac{A_{\perp}}{g_{\perp}} = \frac{A_{J}}{g_{J}}$ with $A_{J}$~=~\SI{-125.3}{\mega\hertz} and the erbium Land\'{e} $g$-factor $g_{J}=6/5$ \cite{AbragamBleaney}. Moreover, the negative sign reproduces the energy level splitting given in Ref.~\cite{Gerasimov2016}.

In \Cref{fig:fig2}(a), the blue dashed lines indicate calculated transition frequencies by taking into account the Hamiltonian parameters from Ref.~\cite{Sattler1971}, which do not include the quadrupole interaction ($P=\SI{0}{\mega\hertz}$). The red dotted lines are the transition frequencies calculated by taking into account the Hamiltonian parameters derived in this work, including the quadrupole interaction ($P= \SI{-7.184}{\mega\hertz}$).
Comparison of the measured zero-field spectrum to the calculated transition frequencies indicates a significantly better matching of the transition frequencies extracted with the spin Hamiltonian parameters of the present study.
The significant impact of the quadrupole interaction on the zero-field spectrum can be clearly seen and is most prominent for those transitions splitting into two peaks, for example, between \SI{2.6}{\giga\hertz} and \SI{2.8}{\giga\hertz}. In general, the fitted spin Hamiltonian parameters $A_{\perp}$, $A_{\parallel}$, and $P$ reproduce the measured transition frequencies well. However, it is also visible that some of the absorption peaks exhibit an additional fine structure. This observation provides evidence that there is an additional interaction, which is not yet included in the Hamiltonian of \Cref{equation:1}. Its origin can be manifold, with likely candidates being the superhyperfine interaction with the nuclear spins of the neighboring fluorine ions, as described for Nd:LiYF$_{4}$ \cite{Macfarlane1998}, the interaction with other paramagnetic impurities, the change of local symmetries, or higher order terms in the Hamiltonian. 
The presence of a fine structure, moreover, implies a larger error of the fitted parameters, as compared to the ones specified from the fitting procedure, due to the uncertainty of the extracted peak frequency.
\subsection{Magnetic field dependence} \label{Sec:Magnetic field}

To further highlight the significance of the quadrupole interaction, we analyze the magnetic field dependence of the hyperfine transitions in the magnetic field range from \SI{0}{\milli\tesla} to \SI{100}{\milli\tesla} for the two previously compared sets of spin Hamiltonian parameters, i.e. those published in Ref.~\cite{Sattler1971} neglecting the quadrupole interaction and the ones determined in this work, which have been used for the calculated spectra shown in \Cref{fig:fig2}(b). 
The discrepancies between the simulated frequencies obtained with the parameters published in Ref.~\cite{Sattler1971} (blue) and those obtained in this work (red) are visible up to approx.~$B_{0}<\SI{50}{\milli\tesla}$. In this field range, the parameters of Ref.~\cite{Sattler1971} suggest degenerate transitions, for example, at approx. \SI{2.7}{\giga\hertz} and \SI{3.1}{\giga\hertz}. However, these transitions become split when taking into account a finite quadrupole interaction, as indicated in the inset in \Cref{fig:fig2}(b). It is evident that above $B_{0}>\SI{50}{\milli\tesla}$ both sets of parameters yield closely coinciding transition frequencies. 
This suggests that in publications neglecting the quadrupole interaction in fitting the data, the derived hyperfine parameters $A_{\parallel}$ and $A_{\perp}$ to some extent contain the quadrupole interaction. Note that with increasing magnetic field, the splitting arising from the quadrupole interaction is no longer resolvable due to the substantial intensity reduction in the higher frequency branch, which makes the identification of the line splitting difficult.

\begin{figure*}
\includegraphics[width=0.99\textwidth]{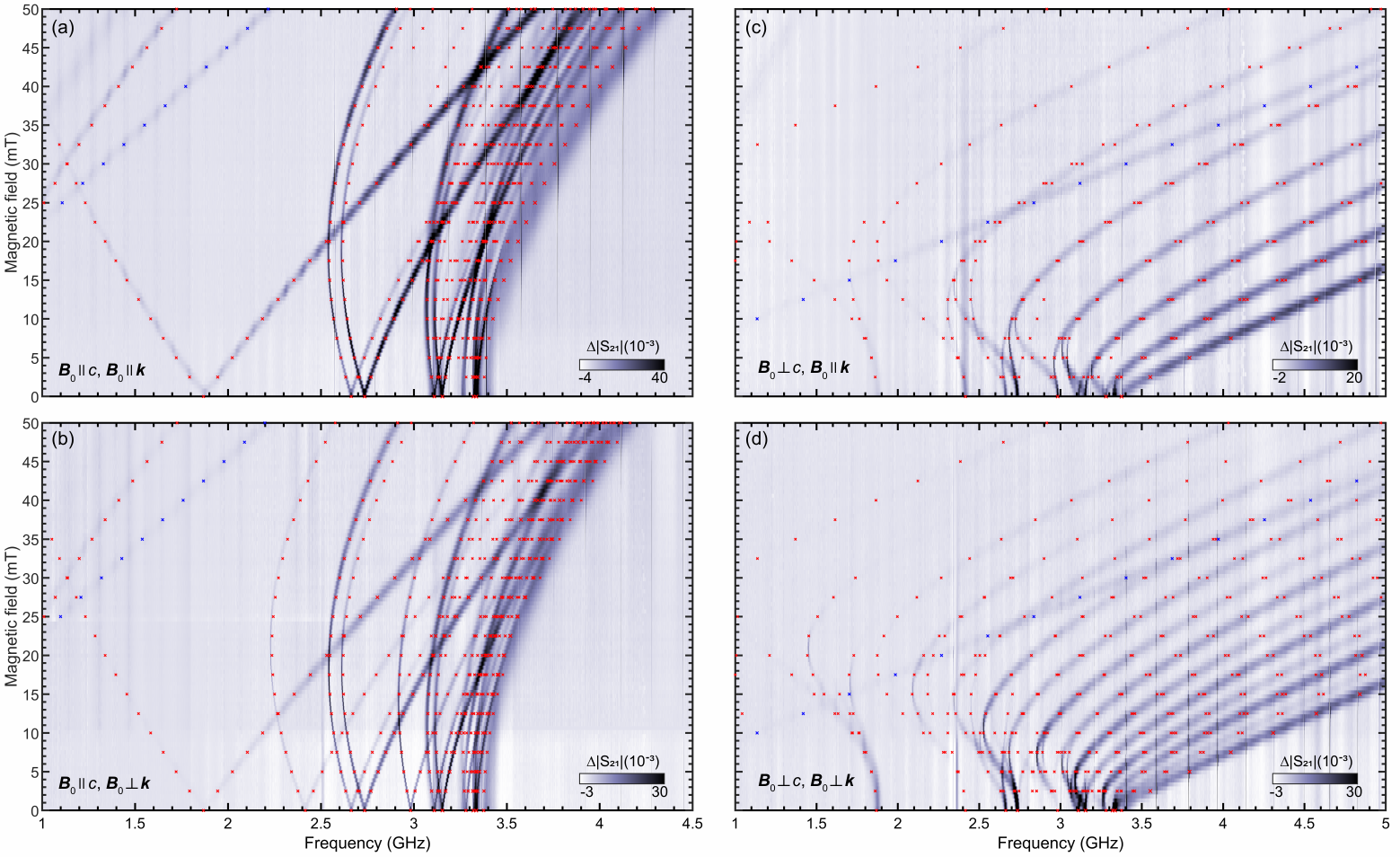}
\caption{The broadband EPR spectra measured in the magnetic field range from \SI{0}{\milli\tesla} to \SI{50}{\milli\tesla} for (a)~$~\boldsymbol{B_{0}} \parallel c$, $\boldsymbol{B_{0}} \parallel \boldsymbol{k}$; (b)~$\boldsymbol{B_{0}} \parallel c$, $\boldsymbol{B_{0}} \perp \boldsymbol{k}$; (c)~$\boldsymbol{B_{0}} \perp c$, $\boldsymbol{B_{0}} \parallel k$; (d)~$\boldsymbol{B_{0}} \perp c$, $\boldsymbol{B_{0}} \perp \boldsymbol{k}$.
The simulated hyperfine ($I = 7/2$) transition frequencies are plotted on top of the measured EPR data with red crosses, while blue crosses represent the erbium-EPR transition for $I = 0$. }
\label{fig:fig3}
\end{figure*}

To obtain the full set of the spin Hamiltonian parameters, we measured broadband EPR spectra in the magnetic field range between \SI{0}{\milli\tesla} and \SI{50}{\milli\tesla} both for $\boldsymbol{B_{0}} \parallel \boldsymbol{c}$ and $\boldsymbol{B_{0}} \perp \boldsymbol{c}$. 
The transmission spectra were taken for four different configurations between external magnetic field, microwave propagation vector $\boldsymbol{k}$, and crystallographic $\boldsymbol{c}$-axis at various fixed magnetic field values separated by field steps of \SI{0.5}{\milli\tesla}. 
Spectra taken in Faraday-like geometry ($\boldsymbol{k} \parallel \boldsymbol{B_{0}}$) are plotted in \Cref{fig:fig3}~(a) for $\boldsymbol{B_{0}} \parallel \boldsymbol{c}$ and in \Cref{fig:fig3}~(c) for $\boldsymbol{B_{0}} \perp \boldsymbol{c}$, while those taken in Voigt-like geometry ($\boldsymbol{k} \perp \boldsymbol{B_{0}}$) are plotted in \Cref{fig:fig3}~(b) for $\boldsymbol{B_{0}} \parallel \boldsymbol{c}$ and in \Cref{fig:fig3}~(d) for $\boldsymbol{B_{0}} \perp \boldsymbol{c}$. Although the studied crystals are doped with 98\% isotopically purified $^{167}$Er, we are able to resolve the $I = 0$ erbium ($^{166,168,170}$Er) EPR-transitions, demonstrating the high sensitivity of the technique. We utilize the $I = 0$ erbium EPR-transition, indicated by the blue crosses in \Cref{fig:fig3}, to confirm the published $g$-factors $g_{\parallel} = 3.137$ and $g_{\perp} = 8.105$ and to extract the misalignment angle $\angle{(B, c)}$ between $\boldsymbol{B_{0}}$ and the $\boldsymbol{c}$-axis. 

For the $\boldsymbol{B_{0}} \parallel \boldsymbol{c}$ configuration, the misalignment angle $\angle{(B, c)}$ was determined to be $3.3^{\circ}$ for $\boldsymbol{B_{0}} \parallel \boldsymbol{c}$, $\boldsymbol{k} \parallel \boldsymbol{B_{0}}$ [\Cref{fig:fig3}~(a)], and $1.3^{\circ}$ for ~$\boldsymbol{B_{0}} \parallel \boldsymbol{c}$, $\boldsymbol{k} \perp \boldsymbol{B_{0}}$ [\Cref{fig:fig3}~(b)]. This information is included in the respective simulations. To account for the varying experimental conditions between the different data sets, we adjust the minimal threshold intensity of the simulated transitions included in the plots in \Cref{fig:fig3}~(a)~-~(d) to represent the experimentally observed transitions. As the utilized background removal technique does not take into account the frequency dependence of the transmission, a quantitative analysis of the extracted intensity and the simulation is not feasible, hence, we restrict the study to a qualitative comparison.

The signal intensity of the measured absorption features depends on the transition probability as given in \Cref{equation:2} as well as on several experimental parameters such as the incident microwave power and the sample temperature. As the sample is placed in an ultra-low temperature environment with the temperature of the mixing chamber stage $T_{\text{MXC}} < \SI{10}{\milli\kelvin}$, the actual temperature of the spin ensemble can differ due to the low thermal conductivity of the sample, a considerable thermal boundary resistance, long spin-lattice-relaxation $T_{1}$, and local heating induced by the probing power \cite{Kukharchyk2020, Kukharchyk2021}.
We estimate the effective temperature of the spin system to $T_{\text{eff}} \approx \SI{20}{\milli\kelvin}$ from the measured temperature dependence of the EPR signal as discussed in \Cref{appendix:c} and include $T_{\text{eff}}$ in our simulations. 

Comparing the spectra in \Cref{fig:fig3}, a different number of allowed transitions is observed for the different excitation geometries. This is expected, since in Voigt-like geometry (see \Cref{fig:fig3}(b) and (d)) excitations in parallel ($\boldsymbol{B_{1}} \parallel \boldsymbol{B_{0}}$) as well as in perpendicular ($\boldsymbol{B_{1}} \perp \boldsymbol{B_{0}}$) mode are present, while in Faraday-like geometry (see \Cref{fig:fig3}(a) and (c)) only perpendicular mode ($\boldsymbol{B_{1}} \perp \boldsymbol{B_{0}}$) excitations are probed. A comparative scheme of the experimental alignment for the two configurations is given in \Cref{appendix:b}. 

Many experimental EPR spectroscopy studies are performed with perpendicular mode excitation, corresponding in our case to a Faraday-like geometry, with typical EPR frequencies in the X-band (\SI{9.5}{\giga\hertz}). Under such experimental conditions, i.e. in the high-field limit, transitions with the selection rules $\Delta m_{S} = \pm 1$ and $\Delta m_{I} = 0$ are forbidden. They can, however, be allowed in the parallel mode configuration if also a state mixing is present, as is the case for the hyperfine levels of the studied spin system of  $^{167}$Er:$^{7}$LiYF$_4$ in the low magnetic field range \cite{Mitrikas2020, Rudowicz2021}.
The different selection rules can be accurately simulated with \Cref{equation:2}, when adapting $\boldsymbol{B_{1}}$ to the respective excitation geometry.
Simulating the transition probability with the general magnetic transition dipole moments defined in Ref.~\cite{Nehrkorn2015} for Voigt and Faraday configuration (for unpolarized light) delivers the same result. Furthermore, analytical solutions given in Ref.~\cite{AbragamBleaney} describe the zero field energy level splitting (for $\boldsymbol{B_{0}} \parallel \boldsymbol{c}$) well, allowing for a representation of the selection rules in terms of the total angular momentum quantum number $m_{F} = m_{S} + m_{I}$, with $\Delta m_{F} = \pm 1$ for $\boldsymbol{B_{1}} \perp \boldsymbol{c}$ and $\Delta m_{F} = \pm 0$ for $\boldsymbol{B_{1}} \parallel \boldsymbol{c}$.
By analyzing the parallel mode excitations, the sign of $P$ relative to $A_{\parallel}$ is determined and found to be negative. A more detailed discussion on the selection rules and the hyperfine states involved is out of the scope of this work. We note, however, that the fitted spin Hamiltonian parameters describe the magnetic field dependence of the measured hyperfine transitions with high precision. 

\subsection{Linewidth analysis}

Besides the frequency of the transition lines, the broadband spectra give direct access to extracting the linewidth of the hyperfine transitions.
The linewidth of the spin transitions is limited by various factors, such as the presence of spatially varying strain within the crystal, the distribution of defects, and fluctuations of the local magnetic field. The magnetic noise can, for example, originate from spin flips of surrounding paramagnetic impurities or the nuclear spin bath of the host crystal, in case it contains ions with $I \neq 0$. 
The inhomogeneous linewidth of optical transitions in isotopically engineered $^{7}$LiYF$_{4}$ was found to be mainly limited by the interaction of the Er$^{3+}$-ions with local magnetic fields originating from fluorine nuclei in the host lattice \cite{Macfarlane1992, Kukharchyk2018}.
Hence, we assume that dipole interactions of this kind dominate the linewidth of the hyperfine transitions also in the microwave range.
Other effects contributing to linewidth broadening, such as charge defects, which are the main linewidth broadening mechanism for Er:CaWO$_{4}$ \cite{Mims1965}, are assumed to be less relevant for $^{167}$Er:$^7$LiYF$_{4}$. The reason is that the substitution of Y$^{3+}$ by Er$^{3+}$ does not require any charge compensation as both ions are in the trivalent state \cite{Chukalina1999}.

\begin{figure}[h]
\includegraphics[width=0.47\textwidth]{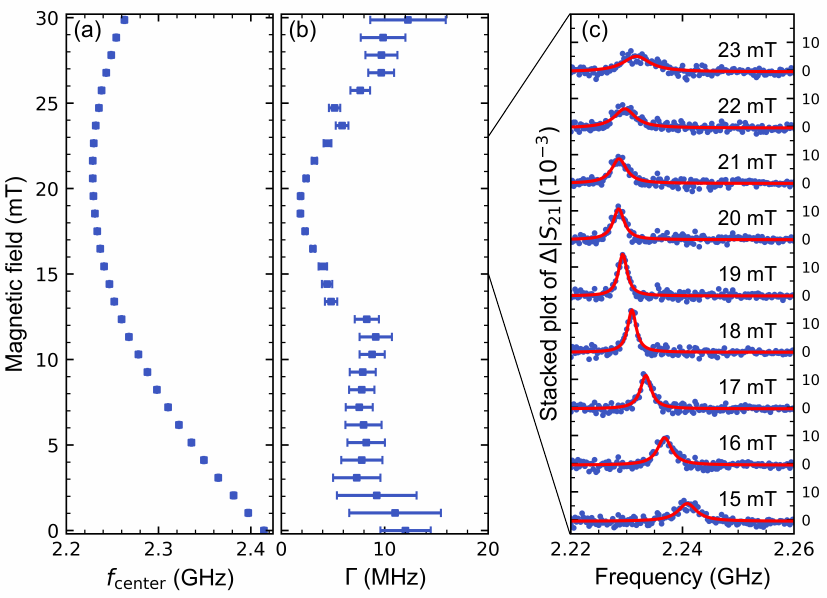}
\caption{Experimentally determined magnetic field dependence of the (a) center frequency $f_{\text{center}}$ and (b) linewidth $\Gamma$ extracted from a Lorentzian fit of $\Delta |S_{21}(f)|$. (c)~Fitting of the absorption line. Stacked plot of the measured $\Delta |S_{21}(f)|$ (blue data points), while the red line indicates the Lorentzian fit at the indicated magnetic fields. The magnitude of $\Delta |S_{21}|$ for each measured trace is indicated on the right.}
\label{fig:fig4}
\end{figure}

We find several transitions that exhibit a narrowing of the linewidth with increasing magnetic field up to the magnetic field value, where the slope of the transition frequency vs. magnetic field dependence vanishes, $\partial f_\text{center}/\partial B_0 \rightarrow 0$. Above this value, the linewidth increases again with increasing field. Due to the plethora of hyperfine transitions exhibiting such behavior, we further exemplary analyze the magnetic field dependence of the linewidth of the $f_{B_{0}=0\,\text{mT}} = \SI{2.415}{\giga\hertz}$ transition for the orientation $\boldsymbol{B_{0}} \parallel \boldsymbol{c}$ and $\boldsymbol{k} \perp \boldsymbol{c}$. This transition was previously identified to exhibit a ZEFOZ behavior at such a turning point at approximately $B_{0} \simeq \SI{20}{\milli\tesla}$~\cite{Kukharchyk2020}. 

The magnetic field dependence of the linewidth $\Gamma$ [extracted as the full width at half maximum (FWHM) of the Lorentzian lineshape, as indicated in the raw data overlayed by the fit in \Cref{fig:fig4}~(c)] of the respective transition in the magnetic field range from $B_{0} = \SI{0}{\milli\tesla}$ to $B_{0} = \SI{30}{\milli\tesla}$, is shown in \Cref{fig:fig4}~(b). The variation of the transition frequency as a function of the applied magnetic field is plotted in \Cref{fig:fig4}~(a). 
We find that $\Gamma$ decreases with increasing magnetic field with a minimal linewidth $\Gamma_{B_{0}\simeq\SI{19}{\milli\tesla}}=\SI{1.84}{\mega\hertz}$. This result reproduces the linewidth dependence reported in \cite{Kukharchyk2020}. While the mentioned electromagnetically induced transparency study extracted the minimal broadening of the spin transition to be $\Gamma_\text{HF0} \simeq 2\pi \times \SI{4.5}{\mega\hertz}$, the substantially lower linewidth in the present study can possibly arise due to the lower spin temperature and smaller magnetic noise of the environment. However, as mentioned in \Cref{appendix:d}, the extracted linewidth is prone to significant uncertainty due to the utilized background removal.
We attribute the observed linewidth narrowing to a reduced sensitivity to magnetic field fluctuations, which is present at the ZEFOZ point \cite{McAuslan2012}.
We further attribute the slight discrepancy between $B_{\Gamma, \min} \simeq \SI{19}{\milli\tesla}$ and the turning point of the transition frequency $B_{S_{1}\rightarrow 0} \simeq \SI{20.5}{\milli\tesla}$ to a finite misalignment of the crystallographic axes of the host crystal with respect to the direction of the applied the magnetic field, since a true ZEFOZ transition is possible only for perfect spatial alignment in all three spatial dimensions \cite{Fraval2004}. 
The importance of a precise alignment of the magnetic field was also pointed out by Fraval et al.~\cite{Fraval2005} for a clock transition in Pr$^{3+}$:Y$_{2}$SiO$_{5}$. As mentioned above, we have determined a misalignment angle of $\angle{(B, c)} \simeq 3.3^{\circ}$, which results in a significant deviation from the conditions for a true ZEFOZ-point. 

\section{Outlook and conclusion}

In conclusion, we have performed broadband EPR spectroscopy on $^{167}$Er:$^{7}$LiYF$_{4}$ single crystals coupled to a superconducting coplanar transmission line at a cryogenic temperature of \SI{10}{\milli\kelvin}. By sweeping the magnetic field in the range from $B_{0} 
=\SI{0}{\milli\tesla}$ to $B_{0}=\SI{50}{\milli\tesla}$ with different orientations of the magnetic field relative to the microwave propagation vector and the crystal $c$-axis, we are able to obtain very precise, refined parameters for the Zeeman, hyperfine, and quadrupole interaction entering the spin Hamiltonian. Most importantly, we find that it is required to take into account the quadrupole interaction for correctly fitting the measured spectra. This is particularly true for the low magnetic field range.

We further analyze the linewidth of the $f_{B_{0}=0\,\text{mT}} = \SI{2.415}{\giga\hertz}$ hyperfine transition around its ZEFOZ-point and find clear evidence for a narrowing of the linewidth down to \SI{1.84}{\mega\hertz}. 
While the detailed origin of an additional fine-structure visible at zero and low magnetic field still remains open, we demonstrate that broadband EPR spectroscopy at ultra-low temperature is a versatile and extremely powerful tool for the investigation and precise characterization of the hyperfine transitions of diluted rare-earth spin ensembles, their selection rules, and linewidth. Due to the strongly increased population difference between the initial and final states, a high signal intensity is obtained, making low-power microwave spectroscopy measurements possible. Moreover, the broadband nature of this technique allows one to directly address various hyperfine transitions simultaneously, potentially at their ZEFOZ point. This makes this technique very promising for microwave quantum memory applications.

\begin{acknowledgments}
We thank Stella Korableva, Oleg Morozov, and Alexey Kalachev for providing the $^{167}$Er:$^{7}$LiYF$_{4}$ crystals and for valuable scientific discussions. We further thank Owen Huisman for the design of the CPW. We acknowledge support by the German Research Foundation via Germany’s Excellence Strategy (EXC-2111- 390814868) and the German Federal Ministry of Education and Research via the project QuaMToMe (Grant No.~16KISQ036).
\end{acknowledgments}

\appendix
\section{Background subtraction} \label{appendix:d}

In our broadband experiments, we removed background contributions related to spectral features of the microwave setup by subtracting a reference background from each measured trace, as described by \Cref{equation:4}. 

We point out that it has been analyzed in Ref.~\cite{Bilzer2007} how different background removal methods affect the extracted resonance frequency and linewidth for the case of ferromagnetic resonance experiments conducted with CPWs.
The simple subtraction approach (as applied in this work) was shown to have a relative error of around 1\% for extracting the resonance frequency and a relative error of up to 10\% for the extraction of the linewidth~\cite{Bilzer2007}.
Therefore, we assume that the transition frequencies extracted in our study are reasonably accurate. However, we also point out that there may be a significant uncertainty in the extracted linewidth.

\section{Calibration of the magnetic field} \label{appendix:a}

To calibrate the strength of the magnetic field, we rely on the EPR transition of 2,2-Diphenyl-1-Picrylhydrazyl (DPPH) sample (Sigma Aldrich), which is placed next to the crystal on top of the superconducting CPW. DPPH is a commonly used magnetic field reference material for EPR spectroscopy studies with an isotropic $g$-factor of 2.0037~\cite{VelluirePellat2023}. To get the magnetic field dependence of the transition frequency, we first fitted the peak corresponding to the EPR transition of DPPH. 

Assuming the known $g$-factor of DPPH, this allows to quantify the magnetic field at the location of the DPPH sample, in close proximity to the $^{167}$Er:$^{7}$LiYF$_4$ sample, as depicted in~\Cref{fig:appendix3}(c). Using this procedure, we obtain a magnetic field calibration for each of the field configurations discussed in the main text, which is crucial for the extraction of exact absolute values of the parameters entering the spin Hamiltonian.
Since the SNR of the DPPH signal is very small, we cannot resolve the published drop of the DPPH $g$-factor as published in earlier studies \cite{VelluirePellat2023}. 
In the case of a too-low signal intensity of the DPPH signal,  we utilize the EPR signal of the erbium isotopes with $I = 0$ to calibrate the magnetic field strength.  

\section{Sample alignment} \label{appendix:b}

\begin{figure}[tbh]
\includegraphics[width=0.47\textwidth]{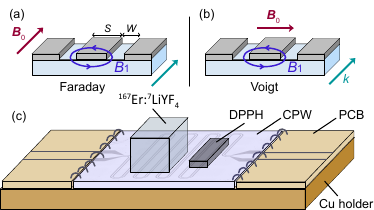}
\caption{Relative orientation of the $\boldsymbol{B_0}$, $\textbf{k}$ and $\boldsymbol{B_{1}}$ vectors for (a) Faraday-like and (b) Voigt-like geometries. (c) Schematic placement of the $^{167}$Er:$^{7}$LiYF$_4$ crystal and the DPPH sample on top of the CPW.}
\label{fig:appendix3}
\end{figure}

The alignment of the applied magnetic field $\boldsymbol{B_0}$, the microwave propagation vector $\boldsymbol{k}$ as well as the oscillating microwave field $\boldsymbol{B_{1}}$ is schematically shown in \Cref{fig:appendix3}~(a)~and~(b) for the Faraday-like and Voigt-like geometries, respectively.
In Faraday-like geometry, the direction of the oscillating microwave magnetic field $\boldsymbol{B_{1}}$ is always perpendicular to the applied static magnetic field, while in Voigt-like geometry, $\boldsymbol{B_{1}}$ has both a parallel and perpendicular component. 
In the CPW design used in our experiments, the width of the central conductor equals $S = \SI{20}{\micro\meter}$, while the width of the gap between the center conductor and the ground is $W = \SI{12}{\micro\meter}$. The placement of the crystal and the DPPH sample on top of the CPW is schematically shown in \Cref{fig:appendix3}~(c). The CPW is wire-bonded to the  printed circuit board (PCB) via the taper patch, while the SMA connectors are then directly connected to the CPW (not shown in \Cref{fig:appendix3}~(c) for clarity). The PCBs and the silicon chip are glued on a copper plate to ensure good thermal coupling to the mixing chamber of the dilution refrigerator.

\section{Temperature dependence} \label{appendix:c}

\begin{figure}[tbh]
\includegraphics[width=0.48\textwidth]{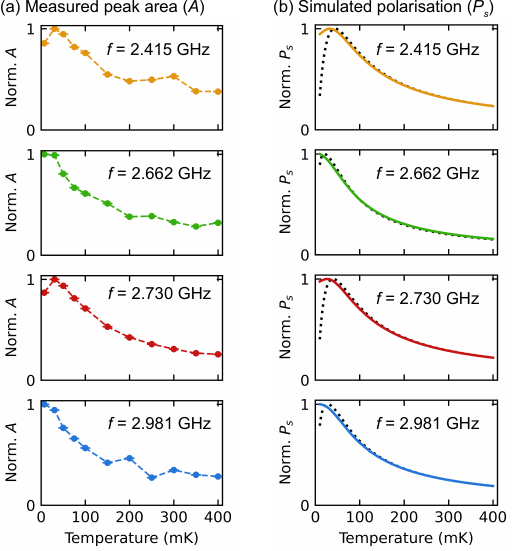}
\caption{(a) Temperature dependence of the experimentally determined peak area ($A$), normalized to its maximum for four absorption peaks with center frequencies $f$= \SI{2.415}{\giga\hertz}, \SI{2.663}{\giga\hertz}, \SI{2.730}{\giga\hertz} and \SI{2.981}{\giga\hertz} at $B_{0}$= \SI{0}{\milli\tesla}. The dashed line serves as a guide to the eye. (b) Simulated temperature dependence of the spin polarisation $P_{s}$ of the transitions respective to (a),  calculated according to \Cref{equation:App3} and normalized to the maximum value. The solid colored line indicates $P_{s}(T_\text{eff})$, while the black dashed line indicates $P_{s}(T)$.}
\label{fig:appendix2}
\end{figure}

A discrepancy between the effective temperature $T_\text{eff}$ of the spin system and the temperature monitored by the sensor placed on the mixing chamber is a well-known phenomenon. Possible reasons are a low thermal conductivity of the sample, local heating by the applied excitation, and a high thermal boundary resistance between sample and heat sink. To account for this discrepancy, we introduce an effective temperature as introduced in Ref.~\cite{Kukharchyk2020, Kukharchyk2021}:
\begin{equation}
    T_\text{eff} = T_\text{min} \left( 1 + \left( T/T_{\text{min}} \right) ^2 \right) ^{1/2} \; .
\label{equation:App1}
\end{equation}
Here, $T_\text{min}$ is the minimal temperature attainable by the spin system, and $T$ indicates the temperature measured by the sensor at the mixing chamber temperature stage. We take broadband spectra at $B_{0} = \SI{0}{\milli\tesla}$ at different temperatures of the mixing chamber stage and derive the temperature dependence of the peak area of various absorption lines with centre frequencies $f$= \SI{2.415}{\giga\hertz}, \SI{2.663}{\giga\hertz}, \SI{2.730}{\giga\hertz} and \SI{2.981}{\giga\hertz}, which are plotted in \Cref{fig:appendix2}~(a). For better comparison, the area of each transition line is normalized to its maximum value. As the change in the peak area relates to the change in the absorption strength, we compare the change in the peak area against the temperature dependence of the spin polarisation of the respective transitions.
To do so, we follow the steps explained in detail in Ref.~\cite{Zollitsch2015}. We first calculate the population probability of each of the 16 hyperfine levels in thermal equilibrium by
\begin{equation}
    p_{t}(T) = \dfrac{e^{\left( -E_{t}/k_\text{B} T \right)}}{\sum_{n=1..16} e^{ \left( -E_{n}/k_\text{B} T \right)}} \; ,
\label{equation:App2}
\end{equation}
with the Boltzmann constant $k_\text{B}$ and temperature $T$, and extract the spin polarisation $P_{s}(T)$ of each studied transition as
\begin{equation}
    P_{s}(T) = |p_{f}(T) - p_{i}(T)| \; ,
\label{equation:App3}
\end{equation}
where $f$ and $i$ indicate the final and initial state. 
To check for the validity of the effective temperature model, we compare the change in peak area to the change in the spin polarisation by computing it in two ways: 
(i) as $P_{s}(T)$ according to \Cref{equation:App3} by utilizing the temperature of the temperature sensor, indicated as the black dashed line in \Cref{fig:appendix2}~(b); 
and (ii) by computing $P_{s}(T_\text{eff})$ according to \Cref{equation:App3} by taking into account $T_\text{eff}$ from \Cref{equation:App1} with $T_\text{min} = \SI{20}{\milli\kelvin}$, plotted in \Cref{fig:appendix2}~(b) with the coloured solid line. For better comparison, $P_{s}(T_\text{eff})$ and $P_{s}(T)$ are normalized to their maximum values, respectively.

The close resemblance between the temperature dependence of $P_{s}(T_\text{eff})$ and the experimentally determined peak area and the divergence when accounting for the sensor temperature highlight the necessity of taking the effective temperature into consideration when working at ultra-low temperatures. Compared to previous studies on the spin temperature for similar samples in the sub-Kelvin temperature range \cite{ Kukharchyk2018,  Kukharchyk2020, Kukharchyk2021, Mladenov2022} with observed $T_\text{min}$ ranging between \SI{70}{\milli\kelvin} and \SI{500}{\milli\kelvin}, the extracted $T_\text{min} \approx T_\text{eff} \approx \SI{20}{\milli\kelvin}$ in this study indicates an efficient cooling of the spin system and a low heating due to the low microwave probing power compared to optical excitation.

In general, the minimal attainable temperature, and thus the effective temperature of the spin ensemble, can vary between separate cooldowns due to differences in sample mounting and thermal anchoring. For our experiments, we conclude that the extracted effective spin temperature gives a good approximation of the spin temperature at the base temperature and is consistent between different data sets, which follows from the similar intensities of the observed peaks in the spectra taken during different cool-downs.

\bibliography{Er_LYF_Broadband_ESR}

\begin{thebibliography}{58}%
\makeatletter
\providecommand \@ifxundefined [1]{%
 \@ifx{#1\undefined}
}%
\providecommand \@ifnum [1]{%
 \ifnum #1\expandafter \@firstoftwo
 \else \expandafter \@secondoftwo
 \fi
}%
\providecommand \@ifx [1]{%
 \ifx #1\expandafter \@firstoftwo
 \else \expandafter \@secondoftwo
 \fi
}%
\providecommand \natexlab [1]{#1}%
\providecommand \enquote  [1]{``#1''}%
\providecommand \bibnamefont  [1]{#1}%
\providecommand \bibfnamefont [1]{#1}%
\providecommand \citenamefont [1]{#1}%
\providecommand \href@noop [0]{\@secondoftwo}%
\providecommand \href [0]{\begingroup \@sanitize@url \@href}%
\providecommand \@href[1]{\@@startlink{#1}\@@href}%
\providecommand \@@href[1]{\endgroup#1\@@endlink}%
\providecommand \@sanitize@url [0]{\catcode `\\12\catcode `\$12\catcode
  `\&12\catcode `\#12\catcode `\^12\catcode `\_12\catcode `\%12\relax}%
\providecommand \@@startlink[1]{}%
\providecommand \@@endlink[0]{}%
\providecommand \url  [0]{\begingroup\@sanitize@url \@url }%
\providecommand \@url [1]{\endgroup\@href {#1}{\urlprefix }}%
\providecommand \urlprefix  [0]{URL }%
\providecommand \Eprint [0]{\href }%
\providecommand \doibase [0]{https://doi.org/}%
\providecommand \selectlanguage [0]{\@gobble}%
\providecommand \bibinfo  [0]{\@secondoftwo}%
\providecommand \bibfield  [0]{\@secondoftwo}%
\providecommand \translation [1]{[#1]}%
\providecommand \BibitemOpen [0]{}%
\providecommand \bibitemStop [0]{}%
\providecommand \bibitemNoStop [0]{.\EOS\space}%
\providecommand \EOS [0]{\spacefactor3000\relax}%
\providecommand \BibitemShut  [1]{\csname bibitem#1\endcsname}%
\let\auto@bib@innerbib\@empty
\bibitem [{\citenamefont {Kurizki}\ \emph {et~al.}(2015)\citenamefont
  {Kurizki}, \citenamefont {Bertet}, \citenamefont {Kubo}, \citenamefont
  {Mølmer}, \citenamefont {Petrosyan}, \citenamefont {Rabl},\ and\
  \citenamefont {Schmiedmayer}}]{Kurizki2015}%
  \BibitemOpen
  \bibfield  {author} {\bibinfo {author} {\bibfnamefont {G.}~\bibnamefont
  {Kurizki}}, \bibinfo {author} {\bibfnamefont {P.}~\bibnamefont {Bertet}},
  \bibinfo {author} {\bibfnamefont {Y.}~\bibnamefont {Kubo}}, \bibinfo {author}
  {\bibfnamefont {K.}~\bibnamefont {Mølmer}}, \bibinfo {author} {\bibfnamefont
  {D.}~\bibnamefont {Petrosyan}}, \bibinfo {author} {\bibfnamefont
  {P.}~\bibnamefont {Rabl}},\ and\ \bibinfo {author} {\bibfnamefont
  {J.}~\bibnamefont {Schmiedmayer}},\ }\bibfield  {title} {\bibinfo {title}
  {{Quantum technologies with hybrid systems}},\ }\href
  {https://doi.org/10.1073/pnas.1419326112} {\bibfield  {journal} {\bibinfo
  {journal} {Proceedings of the National Academy of Sciences}\ }\textbf
  {\bibinfo {volume} {112}},\ \bibinfo {pages} {3866} (\bibinfo {year}
  {2015})}\BibitemShut {NoStop}%
\bibitem [{\citenamefont {Gouzien}\ and\ \citenamefont
  {Sangouard}(2021)}]{Gouzien2021}%
  \BibitemOpen
  \bibfield  {author} {\bibinfo {author} {\bibfnamefont {E.}~\bibnamefont
  {Gouzien}}\ and\ \bibinfo {author} {\bibfnamefont {N.}~\bibnamefont
  {Sangouard}},\ }\bibfield  {title} {\bibinfo {title} {{Factoring 2048-bit RSA
  Integers in 177 Days with 13436 Qubits and a Multimode Memory}},\ }\href
  {https://doi.org/10.1103/PhysRevLett.127.140503} {\bibfield  {journal}
  {\bibinfo  {journal} {Phys. Rev. Lett.}\ }\textbf {\bibinfo {volume} {127}},\
  \bibinfo {pages} {140503} (\bibinfo {year} {2021})}\BibitemShut {NoStop}%
\bibitem [{\citenamefont {Krantz}\ \emph {et~al.}(2019)\citenamefont {Krantz},
  \citenamefont {Kjaergaard}, \citenamefont {Yan}, \citenamefont {Orlando},
  \citenamefont {Gustavsson},\ and\ \citenamefont {Oliver}}]{Krantz2019}%
  \BibitemOpen
  \bibfield  {author} {\bibinfo {author} {\bibfnamefont {P.}~\bibnamefont
  {Krantz}}, \bibinfo {author} {\bibfnamefont {M.}~\bibnamefont {Kjaergaard}},
  \bibinfo {author} {\bibfnamefont {F.}~\bibnamefont {Yan}}, \bibinfo {author}
  {\bibfnamefont {T.~P.}\ \bibnamefont {Orlando}}, \bibinfo {author}
  {\bibfnamefont {S.}~\bibnamefont {Gustavsson}},\ and\ \bibinfo {author}
  {\bibfnamefont {W.~D.}\ \bibnamefont {Oliver}},\ }\bibfield  {title}
  {\bibinfo {title} {{A quantum engineer's guide to superconducting qubits}},\
  }\href {https://doi.org/10.1063/1.5089550} {\bibfield  {journal} {\bibinfo
  {journal} {Applied Physics Reviews}\ }\textbf {\bibinfo {volume} {6}},\
  \bibinfo {pages} {021318} (\bibinfo {year} {2019})}\BibitemShut {NoStop}%
\bibitem [{\citenamefont {Schneider}\ \emph {et~al.}(2019)\citenamefont
  {Schneider}, \citenamefont {Wolz}, \citenamefont {Pfirrmann}, \citenamefont
  {Spiecker}, \citenamefont {Rotzinger}, \citenamefont {Ustinov},\ and\
  \citenamefont {Weides}}]{Schneider2019}%
  \BibitemOpen
  \bibfield  {author} {\bibinfo {author} {\bibfnamefont {A.}~\bibnamefont
  {Schneider}}, \bibinfo {author} {\bibfnamefont {T.}~\bibnamefont {Wolz}},
  \bibinfo {author} {\bibfnamefont {M.}~\bibnamefont {Pfirrmann}}, \bibinfo
  {author} {\bibfnamefont {M.}~\bibnamefont {Spiecker}}, \bibinfo {author}
  {\bibfnamefont {H.}~\bibnamefont {Rotzinger}}, \bibinfo {author}
  {\bibfnamefont {A.~V.}\ \bibnamefont {Ustinov}},\ and\ \bibinfo {author}
  {\bibfnamefont {M.}~\bibnamefont {Weides}},\ }\bibfield  {title} {\bibinfo
  {title} {{Transmon qubit in a magnetic field: Evolution of coherence and
  transition frequency}},\ }\href
  {https://doi.org/10.1103/PhysRevResearch.1.023003} {\bibfield  {journal}
  {\bibinfo  {journal} {Phys. Rev. Res.}\ }\textbf {\bibinfo {volume} {1}},\
  \bibinfo {pages} {023003} (\bibinfo {year} {2019})}\BibitemShut {NoStop}%
\bibitem [{\citenamefont {Kubo}\ \emph {et~al.}(2011)\citenamefont {Kubo},
  \citenamefont {Grezes}, \citenamefont {Dewes}, \citenamefont {Umeda},
  \citenamefont {Isoya}, \citenamefont {Sumiya}, \citenamefont {Morishita},
  \citenamefont {Abe}, \citenamefont {Onoda}, \citenamefont {Ohshima},
  \citenamefont {Jacques}, \citenamefont {Dr\'eau}, \citenamefont {Roch},
  \citenamefont {Diniz}, \citenamefont {Auffeves}, \citenamefont {Vion},
  \citenamefont {Esteve},\ and\ \citenamefont {Bertet}}]{Kubo2011}%
  \BibitemOpen
  \bibfield  {author} {\bibinfo {author} {\bibfnamefont {Y.}~\bibnamefont
  {Kubo}}, \bibinfo {author} {\bibfnamefont {C.}~\bibnamefont {Grezes}},
  \bibinfo {author} {\bibfnamefont {A.}~\bibnamefont {Dewes}}, \bibinfo
  {author} {\bibfnamefont {T.}~\bibnamefont {Umeda}}, \bibinfo {author}
  {\bibfnamefont {J.}~\bibnamefont {Isoya}}, \bibinfo {author} {\bibfnamefont
  {H.}~\bibnamefont {Sumiya}}, \bibinfo {author} {\bibfnamefont
  {N.}~\bibnamefont {Morishita}}, \bibinfo {author} {\bibfnamefont
  {H.}~\bibnamefont {Abe}}, \bibinfo {author} {\bibfnamefont {S.}~\bibnamefont
  {Onoda}}, \bibinfo {author} {\bibfnamefont {T.}~\bibnamefont {Ohshima}},
  \bibinfo {author} {\bibfnamefont {V.}~\bibnamefont {Jacques}}, \bibinfo
  {author} {\bibfnamefont {A.}~\bibnamefont {Dr\'eau}}, \bibinfo {author}
  {\bibfnamefont {J.-F.}\ \bibnamefont {Roch}}, \bibinfo {author}
  {\bibfnamefont {I.}~\bibnamefont {Diniz}}, \bibinfo {author} {\bibfnamefont
  {A.}~\bibnamefont {Auffeves}}, \bibinfo {author} {\bibfnamefont
  {D.}~\bibnamefont {Vion}}, \bibinfo {author} {\bibfnamefont {D.}~\bibnamefont
  {Esteve}},\ and\ \bibinfo {author} {\bibfnamefont {P.}~\bibnamefont
  {Bertet}},\ }\bibfield  {title} {\bibinfo {title} {{Hybrid Quantum Circuit
  with a Superconducting Qubit Coupled to a Spin Ensemble}},\ }\href
  {https://doi.org/10.1103/PhysRevLett.107.220501} {\bibfield  {journal}
  {\bibinfo  {journal} {Phys. Rev. Lett.}\ }\textbf {\bibinfo {volume} {107}},\
  \bibinfo {pages} {220501} (\bibinfo {year} {2011})}\BibitemShut {NoStop}%
\bibitem [{\citenamefont {Grezes}\ \emph {et~al.}(2014)\citenamefont {Grezes},
  \citenamefont {Julsgaard}, \citenamefont {Kubo}, \citenamefont {Stern},
  \citenamefont {Umeda}, \citenamefont {Isoya}, \citenamefont {Sumiya},
  \citenamefont {Abe}, \citenamefont {Onoda}, \citenamefont {Ohshima},
  \citenamefont {Jacques}, \citenamefont {Esteve}, \citenamefont {Vion},
  \citenamefont {Esteve}, \citenamefont {M\o{}lmer},\ and\ \citenamefont
  {Bertet}}]{Grezes2014}%
  \BibitemOpen
  \bibfield  {author} {\bibinfo {author} {\bibfnamefont {C.}~\bibnamefont
  {Grezes}}, \bibinfo {author} {\bibfnamefont {B.}~\bibnamefont {Julsgaard}},
  \bibinfo {author} {\bibfnamefont {Y.}~\bibnamefont {Kubo}}, \bibinfo {author}
  {\bibfnamefont {M.}~\bibnamefont {Stern}}, \bibinfo {author} {\bibfnamefont
  {T.}~\bibnamefont {Umeda}}, \bibinfo {author} {\bibfnamefont
  {J.}~\bibnamefont {Isoya}}, \bibinfo {author} {\bibfnamefont
  {H.}~\bibnamefont {Sumiya}}, \bibinfo {author} {\bibfnamefont
  {H.}~\bibnamefont {Abe}}, \bibinfo {author} {\bibfnamefont {S.}~\bibnamefont
  {Onoda}}, \bibinfo {author} {\bibfnamefont {T.}~\bibnamefont {Ohshima}},
  \bibinfo {author} {\bibfnamefont {V.}~\bibnamefont {Jacques}}, \bibinfo
  {author} {\bibfnamefont {J.}~\bibnamefont {Esteve}}, \bibinfo {author}
  {\bibfnamefont {D.}~\bibnamefont {Vion}}, \bibinfo {author} {\bibfnamefont
  {D.}~\bibnamefont {Esteve}}, \bibinfo {author} {\bibfnamefont
  {K.}~\bibnamefont {M\o{}lmer}},\ and\ \bibinfo {author} {\bibfnamefont
  {P.}~\bibnamefont {Bertet}},\ }\bibfield  {title} {\bibinfo {title}
  {{Multimode Storage and Retrieval of Microwave Fields in a Spin Ensemble}},\
  }\href {https://doi.org/10.1103/PhysRevX.4.021049} {\bibfield  {journal}
  {\bibinfo  {journal} {Phys. Rev. X}\ }\textbf {\bibinfo {volume} {4}},\
  \bibinfo {pages} {021049} (\bibinfo {year} {2014})}\BibitemShut {NoStop}%
\bibitem [{\citenamefont {O'Sullivan}\ \emph {et~al.}(2020)\citenamefont
  {O'Sullivan}, \citenamefont {Kennedy}, \citenamefont {Zollitsch},
  \citenamefont {\v{S}im\.{e}nas}, \citenamefont {Thomas}, \citenamefont
  {Abdurakhimov}, \citenamefont {Withington},\ and\ \citenamefont
  {Morton}}]{OSullivan2020}%
  \BibitemOpen
  \bibfield  {author} {\bibinfo {author} {\bibfnamefont {J.}~\bibnamefont
  {O'Sullivan}}, \bibinfo {author} {\bibfnamefont {O.~W.}\ \bibnamefont
  {Kennedy}}, \bibinfo {author} {\bibfnamefont {C.~W.}\ \bibnamefont
  {Zollitsch}}, \bibinfo {author} {\bibfnamefont {M.}~\bibnamefont
  {\v{S}im\.{e}nas}}, \bibinfo {author} {\bibfnamefont {C.~N.}\ \bibnamefont
  {Thomas}}, \bibinfo {author} {\bibfnamefont {L.~V.}\ \bibnamefont
  {Abdurakhimov}}, \bibinfo {author} {\bibfnamefont {S.}~\bibnamefont
  {Withington}},\ and\ \bibinfo {author} {\bibfnamefont {J.~J.}\ \bibnamefont
  {Morton}},\ }\bibfield  {title} {\bibinfo {title} {{Spin-Resonance Linewidths
  of Bismuth Donors in Silicon Coupled to Planar Microresonators}},\ }\href
  {https://doi.org/10.1103/PhysRevApplied.14.064050} {\bibfield  {journal}
  {\bibinfo  {journal} {Phys. Rev. Appl.}\ }\textbf {\bibinfo {volume} {14}},\
  \bibinfo {pages} {064050} (\bibinfo {year} {2020})}\BibitemShut {NoStop}%
\bibitem [{\citenamefont {Zollitsch}\ \emph {et~al.}(2015)\citenamefont
  {Zollitsch}, \citenamefont {Mueller}, \citenamefont {Franke}, \citenamefont
  {Goennenwein}, \citenamefont {Brandt}, \citenamefont {Gross},\ and\
  \citenamefont {Huebl}}]{Zollitsch2015}%
  \BibitemOpen
  \bibfield  {author} {\bibinfo {author} {\bibfnamefont {C.~W.}\ \bibnamefont
  {Zollitsch}}, \bibinfo {author} {\bibfnamefont {K.}~\bibnamefont {Mueller}},
  \bibinfo {author} {\bibfnamefont {D.~P.}\ \bibnamefont {Franke}}, \bibinfo
  {author} {\bibfnamefont {S.~T.~B.}\ \bibnamefont {Goennenwein}}, \bibinfo
  {author} {\bibfnamefont {M.~S.}\ \bibnamefont {Brandt}}, \bibinfo {author}
  {\bibfnamefont {R.}~\bibnamefont {Gross}},\ and\ \bibinfo {author}
  {\bibfnamefont {H.}~\bibnamefont {Huebl}},\ }\bibfield  {title} {\bibinfo
  {title} {High cooperativity coupling between a phosphorus donor spin ensemble
  and a superconducting microwave resonator},\ }\href
  {https://doi.org/10.1063/1.4932658} {\bibfield  {journal} {\bibinfo
  {journal} {Applied Physics Letters}\ }\textbf {\bibinfo {volume} {107}},\
  \bibinfo {pages} {142105} (\bibinfo {year} {2015})}\BibitemShut {NoStop}%
\bibitem [{\citenamefont {Weichselbaumer}\ \emph {et~al.}(2020)\citenamefont
  {Weichselbaumer}, \citenamefont {Zens}, \citenamefont {Zollitsch},
  \citenamefont {Brandt}, \citenamefont {Rotter}, \citenamefont {Gross},\ and\
  \citenamefont {Huebl}}]{Weichselbaumer2020}%
  \BibitemOpen
  \bibfield  {author} {\bibinfo {author} {\bibfnamefont {S.}~\bibnamefont
  {Weichselbaumer}}, \bibinfo {author} {\bibfnamefont {M.}~\bibnamefont
  {Zens}}, \bibinfo {author} {\bibfnamefont {C.~W.}\ \bibnamefont {Zollitsch}},
  \bibinfo {author} {\bibfnamefont {M.~S.}\ \bibnamefont {Brandt}}, \bibinfo
  {author} {\bibfnamefont {S.}~\bibnamefont {Rotter}}, \bibinfo {author}
  {\bibfnamefont {R.}~\bibnamefont {Gross}},\ and\ \bibinfo {author}
  {\bibfnamefont {H.}~\bibnamefont {Huebl}},\ }\bibfield  {title} {\bibinfo
  {title} {Echo trains in pulsed electron spin resonance of a strongly coupled
  spin ensemble},\ }\href {https://doi.org/10.1103/PhysRevLett.125.137701}
  {\bibfield  {journal} {\bibinfo  {journal} {Phys. Rev. Lett.}\ }\textbf
  {\bibinfo {volume} {125}},\ \bibinfo {pages} {137701} (\bibinfo {year}
  {2020})}\BibitemShut {NoStop}%
\bibitem [{\citenamefont {Wolfowicz}\ \emph {et~al.}(2015)\citenamefont
  {Wolfowicz}, \citenamefont {Maier-Flaig}, \citenamefont {Marino},
  \citenamefont {Ferrier}, \citenamefont {Vezin}, \citenamefont {Morton},\ and\
  \citenamefont {Goldner}}]{Wolfowicz2015}%
  \BibitemOpen
  \bibfield  {author} {\bibinfo {author} {\bibfnamefont {G.}~\bibnamefont
  {Wolfowicz}}, \bibinfo {author} {\bibfnamefont {H.}~\bibnamefont
  {Maier-Flaig}}, \bibinfo {author} {\bibfnamefont {R.}~\bibnamefont {Marino}},
  \bibinfo {author} {\bibfnamefont {A.}~\bibnamefont {Ferrier}}, \bibinfo
  {author} {\bibfnamefont {H.}~\bibnamefont {Vezin}}, \bibinfo {author}
  {\bibfnamefont {J.~J.~L.}\ \bibnamefont {Morton}},\ and\ \bibinfo {author}
  {\bibfnamefont {P.}~\bibnamefont {Goldner}},\ }\bibfield  {title} {\bibinfo
  {title} {{Coherent Storage of Microwave Excitations in Rare-Earth Nuclear
  Spins}},\ }\href {https://doi.org/10.1103/PhysRevLett.114.170503} {\bibfield
  {journal} {\bibinfo  {journal} {Phys. Rev. Lett.}\ }\textbf {\bibinfo
  {volume} {114}},\ \bibinfo {pages} {170503} (\bibinfo {year}
  {2015})}\BibitemShut {NoStop}%
\bibitem [{\citenamefont {Afzelius}\ \emph {et~al.}(2013)\citenamefont
  {Afzelius}, \citenamefont {Sangouard}, \citenamefont {Johansson},
  \citenamefont {Staudt},\ and\ \citenamefont {Wilson}}]{Afzelius2013}%
  \BibitemOpen
  \bibfield  {author} {\bibinfo {author} {\bibfnamefont {M.}~\bibnamefont
  {Afzelius}}, \bibinfo {author} {\bibfnamefont {N.}~\bibnamefont {Sangouard}},
  \bibinfo {author} {\bibfnamefont {G.}~\bibnamefont {Johansson}}, \bibinfo
  {author} {\bibfnamefont {M.~U.}\ \bibnamefont {Staudt}},\ and\ \bibinfo
  {author} {\bibfnamefont {C.~M.}\ \bibnamefont {Wilson}},\ }\bibfield  {title}
  {\bibinfo {title} {{Proposal for a coherent quantum memory for propagating
  microwave photons}},\ }\href {https://doi.org/10.1088/1367-2630/15/6/065008}
  {\bibfield  {journal} {\bibinfo  {journal} {New Journal of Physics}\ }\textbf
  {\bibinfo {volume} {15}},\ \bibinfo {pages} {065008} (\bibinfo {year}
  {2013})}\BibitemShut {NoStop}%
\bibitem [{\citenamefont {Probst}\ \emph {et~al.}(2015)\citenamefont {Probst},
  \citenamefont {Rotzinger}, \citenamefont {Ustinov},\ and\ \citenamefont
  {Bushev}}]{Probst2015}%
  \BibitemOpen
  \bibfield  {author} {\bibinfo {author} {\bibfnamefont {S.}~\bibnamefont
  {Probst}}, \bibinfo {author} {\bibfnamefont {H.}~\bibnamefont {Rotzinger}},
  \bibinfo {author} {\bibfnamefont {A.~V.}\ \bibnamefont {Ustinov}},\ and\
  \bibinfo {author} {\bibfnamefont {P.~A.}\ \bibnamefont {Bushev}},\ }\bibfield
   {title} {\bibinfo {title} {{Microwave multimode memory with an erbium spin
  ensemble}},\ }\href {https://doi.org/10.1103/PhysRevB.92.014421} {\bibfield
  {journal} {\bibinfo  {journal} {Phys. Rev. B}\ }\textbf {\bibinfo {volume}
  {92}},\ \bibinfo {pages} {014421} (\bibinfo {year} {2015})}\BibitemShut
  {NoStop}%
\bibitem [{\citenamefont {{Le Dantec}}\ \emph {et~al.}(2021)\citenamefont {{Le
  Dantec}}, \citenamefont {Rančić}, \citenamefont {Lin}, \citenamefont
  {Billaud}, \citenamefont {Ranjan}, \citenamefont {Flanigan}, \citenamefont
  {Bertaina}, \citenamefont {Chanelière}, \citenamefont {Goldner},
  \citenamefont {Erb}, \citenamefont {Liu}, \citenamefont {Estève},
  \citenamefont {Vion}, \citenamefont {Flurin},\ and\ \citenamefont
  {Bertet}}]{LeDantec2021}%
  \BibitemOpen
  \bibfield  {author} {\bibinfo {author} {\bibfnamefont {M.}~\bibnamefont {{Le
  Dantec}}}, \bibinfo {author} {\bibfnamefont {M.}~\bibnamefont {Rančić}},
  \bibinfo {author} {\bibfnamefont {S.}~\bibnamefont {Lin}}, \bibinfo {author}
  {\bibfnamefont {E.}~\bibnamefont {Billaud}}, \bibinfo {author} {\bibfnamefont
  {V.}~\bibnamefont {Ranjan}}, \bibinfo {author} {\bibfnamefont
  {D.}~\bibnamefont {Flanigan}}, \bibinfo {author} {\bibfnamefont
  {S.}~\bibnamefont {Bertaina}}, \bibinfo {author} {\bibfnamefont
  {T.}~\bibnamefont {Chanelière}}, \bibinfo {author} {\bibfnamefont
  {P.}~\bibnamefont {Goldner}}, \bibinfo {author} {\bibfnamefont
  {A.}~\bibnamefont {Erb}}, \bibinfo {author} {\bibfnamefont {R.~B.}\
  \bibnamefont {Liu}}, \bibinfo {author} {\bibfnamefont {D.}~\bibnamefont
  {Estève}}, \bibinfo {author} {\bibfnamefont {D.}~\bibnamefont {Vion}},
  \bibinfo {author} {\bibfnamefont {E.}~\bibnamefont {Flurin}},\ and\ \bibinfo
  {author} {\bibfnamefont {P.}~\bibnamefont {Bertet}},\ }\bibfield  {title}
  {\bibinfo {title} {{Twenty-three–millisecond electron spin coherence of
  erbium ions in a natural-abundance crystal}},\ }\href
  {https://doi.org/10.1126/sciadv.abj9786} {\bibfield  {journal} {\bibinfo
  {journal} {Science Advances}\ }\textbf {\bibinfo {volume} {7}},\ \bibinfo
  {pages} {eabj9786} (\bibinfo {year} {2021})}\BibitemShut {NoStop}%
\bibitem [{\citenamefont {Merkel}\ \emph {et~al.}(2021)\citenamefont {Merkel},
  \citenamefont {Cova Fari\~na},\ and\ \citenamefont {Reiserer}}]{Merkel2021}%
  \BibitemOpen
  \bibfield  {author} {\bibinfo {author} {\bibfnamefont {B.}~\bibnamefont
  {Merkel}}, \bibinfo {author} {\bibfnamefont {P.}~\bibnamefont {Cova
  Fari\~na}},\ and\ \bibinfo {author} {\bibfnamefont {A.}~\bibnamefont
  {Reiserer}},\ }\bibfield  {title} {\bibinfo {title} {{Dynamical Decoupling of
  Spin Ensembles with Strong Anisotropic Interactions}},\ }\href
  {https://doi.org/10.1103/PhysRevLett.127.030501} {\bibfield  {journal}
  {\bibinfo  {journal} {Phys. Rev. Lett.}\ }\textbf {\bibinfo {volume} {127}},\
  \bibinfo {pages} {030501} (\bibinfo {year} {2021})}\BibitemShut {NoStop}%
\bibitem [{\citenamefont {Wolfowicz}\ \emph {et~al.}(2013)\citenamefont
  {Wolfowicz}, \citenamefont {Tyryshkin}, \citenamefont {George}, \citenamefont
  {Riemann}, \citenamefont {Abrosimov}, \citenamefont {Becker}, \citenamefont
  {Pohl}, \citenamefont {Thewalt}, \citenamefont {Lyon},\ and\ \citenamefont
  {Morton}}]{Wolfowicz2013}%
  \BibitemOpen
  \bibfield  {author} {\bibinfo {author} {\bibfnamefont {G.}~\bibnamefont
  {Wolfowicz}}, \bibinfo {author} {\bibfnamefont {A.~M.}\ \bibnamefont
  {Tyryshkin}}, \bibinfo {author} {\bibfnamefont {R.~E.}\ \bibnamefont
  {George}}, \bibinfo {author} {\bibfnamefont {H.}~\bibnamefont {Riemann}},
  \bibinfo {author} {\bibfnamefont {N.~V.}\ \bibnamefont {Abrosimov}}, \bibinfo
  {author} {\bibfnamefont {P.}~\bibnamefont {Becker}}, \bibinfo {author}
  {\bibfnamefont {H.-J.}\ \bibnamefont {Pohl}}, \bibinfo {author}
  {\bibfnamefont {M.~L.~W.}\ \bibnamefont {Thewalt}}, \bibinfo {author}
  {\bibfnamefont {S.~A.}\ \bibnamefont {Lyon}},\ and\ \bibinfo {author}
  {\bibfnamefont {J.~J.~L.}\ \bibnamefont {Morton}},\ }\bibfield  {title}
  {\bibinfo {title} {{Atomic clock transitions in silicon-based spin qubits}},\
  }\href {https://doi.org/10.1038/nnano.2013.117} {\bibfield  {journal}
  {\bibinfo  {journal} {Nature Nanotechnology}\ }\textbf {\bibinfo {volume}
  {8}},\ \bibinfo {pages} {561} (\bibinfo {year} {2013})}\BibitemShut {NoStop}%
\bibitem [{\citenamefont {Ortu}\ \emph {et~al.}(2018)\citenamefont {Ortu},
  \citenamefont {Tiranov}, \citenamefont {Welinski}, \citenamefont {Fröwis},
  \citenamefont {Gisin}, \citenamefont {Ferrier}, \citenamefont {Goldner},\
  and\ \citenamefont {Afzelius}}]{Ortu2018}%
  \BibitemOpen
  \bibfield  {author} {\bibinfo {author} {\bibfnamefont {A.}~\bibnamefont
  {Ortu}}, \bibinfo {author} {\bibfnamefont {A.}~\bibnamefont {Tiranov}},
  \bibinfo {author} {\bibfnamefont {S.}~\bibnamefont {Welinski}}, \bibinfo
  {author} {\bibfnamefont {F.}~\bibnamefont {Fröwis}}, \bibinfo {author}
  {\bibfnamefont {N.}~\bibnamefont {Gisin}}, \bibinfo {author} {\bibfnamefont
  {A.}~\bibnamefont {Ferrier}}, \bibinfo {author} {\bibfnamefont
  {P.}~\bibnamefont {Goldner}},\ and\ \bibinfo {author} {\bibfnamefont
  {M.}~\bibnamefont {Afzelius}},\ }\bibfield  {title} {\bibinfo {title}
  {{Simultaneous coherence enhancement of optical and microwave transitions in
  solid-state electronic spins}},\ }\href
  {https://doi.org/10.1038/s41563-018-0138-x} {\bibfield  {journal} {\bibinfo
  {journal} {Nature Materials}\ }\textbf {\bibinfo {volume} {17}},\ \bibinfo
  {pages} {671–675} (\bibinfo {year} {2018})}\BibitemShut {NoStop}%
\bibitem [{\citenamefont {McAuslan}\ \emph {et~al.}(2012)\citenamefont
  {McAuslan}, \citenamefont {Bartholomew}, \citenamefont {Sellars},\ and\
  \citenamefont {Longdell}}]{McAuslan2012}%
  \BibitemOpen
  \bibfield  {author} {\bibinfo {author} {\bibfnamefont {D.~L.}\ \bibnamefont
  {McAuslan}}, \bibinfo {author} {\bibfnamefont {J.~G.}\ \bibnamefont
  {Bartholomew}}, \bibinfo {author} {\bibfnamefont {M.~J.}\ \bibnamefont
  {Sellars}},\ and\ \bibinfo {author} {\bibfnamefont {J.~J.}\ \bibnamefont
  {Longdell}},\ }\bibfield  {title} {\bibinfo {title} {{Reducing decoherence in
  optical and spin transitions in rare-earth-metal-ion--doped materials}},\
  }\href {https://doi.org/10.1103/PhysRevA.85.032339} {\bibfield  {journal}
  {\bibinfo  {journal} {Phys. Rev. A}\ }\textbf {\bibinfo {volume} {85}},\
  \bibinfo {pages} {032339} (\bibinfo {year} {2012})}\BibitemShut {NoStop}%
\bibitem [{\citenamefont {Rakonjac}\ \emph {et~al.}(2020)\citenamefont
  {Rakonjac}, \citenamefont {Chen}, \citenamefont {Horvath},\ and\
  \citenamefont {Longdell}}]{Rakonjac2020}%
  \BibitemOpen
  \bibfield  {author} {\bibinfo {author} {\bibfnamefont {J.~V.}\ \bibnamefont
  {Rakonjac}}, \bibinfo {author} {\bibfnamefont {Y.-H.}\ \bibnamefont {Chen}},
  \bibinfo {author} {\bibfnamefont {S.~P.}\ \bibnamefont {Horvath}},\ and\
  \bibinfo {author} {\bibfnamefont {J.~J.}\ \bibnamefont {Longdell}},\
  }\bibfield  {title} {\bibinfo {title} {{Long spin coherence times in the
  ground state and in an optically excited state of
  $^{167}\mathrm{Er}^{3+}\!:\!\mathrm{Y}_{2}\mathrm{SiO}_{5}$ at zero magnetic
  field}},\ }\href {https://doi.org/10.1103/PhysRevB.101.184430} {\bibfield
  {journal} {\bibinfo  {journal} {Phys. Rev. B}\ }\textbf {\bibinfo {volume}
  {101}},\ \bibinfo {pages} {184430} (\bibinfo {year} {2020})}\BibitemShut
  {NoStop}%
\bibitem [{\citenamefont {Alizadeh}\ \emph {et~al.}(2023)\citenamefont
  {Alizadeh}, \citenamefont {Wells}, \citenamefont {Reid}, \citenamefont
  {Ferrier},\ and\ \citenamefont {Goldner}}]{Alizadeh2023}%
  \BibitemOpen
  \bibfield  {author} {\bibinfo {author} {\bibfnamefont {Y.}~\bibnamefont
  {Alizadeh}}, \bibinfo {author} {\bibfnamefont {J.-P.~R.}\ \bibnamefont
  {Wells}}, \bibinfo {author} {\bibfnamefont {M.~F.}\ \bibnamefont {Reid}},
  \bibinfo {author} {\bibfnamefont {A.}~\bibnamefont {Ferrier}},\ and\ \bibinfo
  {author} {\bibfnamefont {P.}~\bibnamefont {Goldner}},\ }\bibfield  {title}
  {\bibinfo {title} {{Zeeman spectroscopy and crystal-field analysis of low
  symmetry centres in $\mathrm{Nd}^{3+}$ doped
  $\mathrm{Y}_{2}\mathrm{SiO}_{5}$}},\ }\href
  {https://doi.org/10.1088/1361-648X/acceed} {\bibfield  {journal} {\bibinfo
  {journal} {Journal of Physics: Condensed Matter}\ }\textbf {\bibinfo {volume}
  {35}},\ \bibinfo {pages} {305502} (\bibinfo {year} {2023})}\BibitemShut
  {NoStop}%
\bibitem [{\citenamefont {Clauss}\ \emph {et~al.}(2013)\citenamefont {Clauss},
  \citenamefont {Bothner}, \citenamefont {Koelle}, \citenamefont {Kleiner},
  \citenamefont {Bogani}, \citenamefont {Scheffler},\ and\ \citenamefont
  {Dressel}}]{Clauss2013}%
  \BibitemOpen
  \bibfield  {author} {\bibinfo {author} {\bibfnamefont {C.}~\bibnamefont
  {Clauss}}, \bibinfo {author} {\bibfnamefont {D.}~\bibnamefont {Bothner}},
  \bibinfo {author} {\bibfnamefont {D.}~\bibnamefont {Koelle}}, \bibinfo
  {author} {\bibfnamefont {R.}~\bibnamefont {Kleiner}}, \bibinfo {author}
  {\bibfnamefont {L.}~\bibnamefont {Bogani}}, \bibinfo {author} {\bibfnamefont
  {M.}~\bibnamefont {Scheffler}},\ and\ \bibinfo {author} {\bibfnamefont
  {M.}~\bibnamefont {Dressel}},\ }\bibfield  {title} {\bibinfo {title}
  {{Broadband electron spin resonance from 500 MHz to 40 GHz using
  superconducting coplanar waveguides}},\ }\href
  {https://doi.org/10.1063/1.4802956} {\bibfield  {journal} {\bibinfo
  {journal} {Applied Physics Letters}\ }\textbf {\bibinfo {volume} {102}},\
  \bibinfo {pages} {162601} (\bibinfo {year} {2013})}\BibitemShut {NoStop}%
\bibitem [{\citenamefont {Prinz-Zwick}\ \emph {et~al.}(2022)\citenamefont
  {Prinz-Zwick}, \citenamefont {Szigeti}, \citenamefont {Gimpel}, \citenamefont
  {Ehlers}, \citenamefont {Tsurkan}, \citenamefont {Leonov}, \citenamefont
  {Miksch}, \citenamefont {Scheffler}, \citenamefont {Stasinopoulos},
  \citenamefont {Grundler}, \citenamefont {Kézsmárki}, \citenamefont
  {Büttgen},\ and\ \citenamefont {Krug~von Nidda}}]{Prinz-Zwick2022}%
  \BibitemOpen
  \bibfield  {author} {\bibinfo {author} {\bibfnamefont {M.}~\bibnamefont
  {Prinz-Zwick}}, \bibinfo {author} {\bibfnamefont {B.~G.}\ \bibnamefont
  {Szigeti}}, \bibinfo {author} {\bibfnamefont {T.}~\bibnamefont {Gimpel}},
  \bibinfo {author} {\bibfnamefont {D.}~\bibnamefont {Ehlers}}, \bibinfo
  {author} {\bibfnamefont {V.}~\bibnamefont {Tsurkan}}, \bibinfo {author}
  {\bibfnamefont {A.~O.}\ \bibnamefont {Leonov}}, \bibinfo {author}
  {\bibfnamefont {B.}~\bibnamefont {Miksch}}, \bibinfo {author} {\bibfnamefont
  {M.}~\bibnamefont {Scheffler}}, \bibinfo {author} {\bibfnamefont
  {I.}~\bibnamefont {Stasinopoulos}}, \bibinfo {author} {\bibfnamefont
  {D.}~\bibnamefont {Grundler}}, \bibinfo {author} {\bibfnamefont
  {I.}~\bibnamefont {Kézsmárki}}, \bibinfo {author} {\bibfnamefont
  {N.}~\bibnamefont {Büttgen}},\ and\ \bibinfo {author} {\bibfnamefont
  {H.-A.}\ \bibnamefont {Krug~von Nidda}},\ }\bibfield  {title} {\bibinfo
  {title} {{Nuclear and Electron Spin Resonance Studies on Skyrmion-Hosting
  Lacunar Spinels}},\ }\href
  {https://doi.org/https://doi.org/10.1002/pssb.202100170} {\bibfield
  {journal} {\bibinfo  {journal} {physica status solidi (b)}\ }\textbf
  {\bibinfo {volume} {259}},\ \bibinfo {pages} {2100170} (\bibinfo {year}
  {2022})}\BibitemShut {NoStop}%
\bibitem [{\citenamefont {Wiemann}\ \emph {et~al.}(2015)\citenamefont
  {Wiemann}, \citenamefont {Simmendinger}, \citenamefont {Clauss},
  \citenamefont {Bogani}, \citenamefont {Bothner}, \citenamefont {Koelle},
  \citenamefont {Kleiner}, \citenamefont {Dressel},\ and\ \citenamefont
  {Scheffler}}]{Wiemann2015}%
  \BibitemOpen
  \bibfield  {author} {\bibinfo {author} {\bibfnamefont {Y.}~\bibnamefont
  {Wiemann}}, \bibinfo {author} {\bibfnamefont {J.}~\bibnamefont
  {Simmendinger}}, \bibinfo {author} {\bibfnamefont {C.}~\bibnamefont
  {Clauss}}, \bibinfo {author} {\bibfnamefont {L.}~\bibnamefont {Bogani}},
  \bibinfo {author} {\bibfnamefont {D.}~\bibnamefont {Bothner}}, \bibinfo
  {author} {\bibfnamefont {D.}~\bibnamefont {Koelle}}, \bibinfo {author}
  {\bibfnamefont {R.}~\bibnamefont {Kleiner}}, \bibinfo {author} {\bibfnamefont
  {M.}~\bibnamefont {Dressel}},\ and\ \bibinfo {author} {\bibfnamefont
  {M.}~\bibnamefont {Scheffler}},\ }\bibfield  {title} {\bibinfo {title}
  {{Observing electron spin resonance between 0.1 and 67 GHz at temperatures
  between 50 mK and 300 K using broadband metallic coplanar waveguides}},\
  }\href {https://doi.org/10.1063/1.4921231} {\bibfield  {journal} {\bibinfo
  {journal} {Applied Physics Letters}\ }\textbf {\bibinfo {volume} {106}},\
  \bibinfo {pages} {193505} (\bibinfo {year} {2015})}\BibitemShut {NoStop}%
\bibitem [{\citenamefont {Jing}\ \emph {et~al.}(2019)\citenamefont {Jing},
  \citenamefont {Lan}, \citenamefont {Shi}, \citenamefont {Mu}, \citenamefont
  {Qin}, \citenamefont {Rong},\ and\ \citenamefont {Du}}]{Jing2019}%
  \BibitemOpen
  \bibfield  {author} {\bibinfo {author} {\bibfnamefont {K.}~\bibnamefont
  {Jing}}, \bibinfo {author} {\bibfnamefont {Z.}~\bibnamefont {Lan}}, \bibinfo
  {author} {\bibfnamefont {Z.}~\bibnamefont {Shi}}, \bibinfo {author}
  {\bibfnamefont {S.}~\bibnamefont {Mu}}, \bibinfo {author} {\bibfnamefont
  {X.}~\bibnamefont {Qin}}, \bibinfo {author} {\bibfnamefont {X.}~\bibnamefont
  {Rong}},\ and\ \bibinfo {author} {\bibfnamefont {J.}~\bibnamefont {Du}},\
  }\bibfield  {title} {\bibinfo {title} {{Broadband electron paramagnetic
  resonance spectrometer from 1 to 15 GHz using metallic coplanar waveguide}},\
  }\href {https://doi.org/10.1063/1.5119333} {\bibfield  {journal} {\bibinfo
  {journal} {Review of Scientific Instruments}\ }\textbf {\bibinfo {volume}
  {90}},\ \bibinfo {pages} {125109} (\bibinfo {year} {2019})}\BibitemShut
  {NoStop}%
\bibitem [{\citenamefont {Miksch}\ \emph {et~al.}(2020)\citenamefont {Miksch},
  \citenamefont {Dressel},\ and\ \citenamefont {Scheffler}}]{Miksch2020}%
  \BibitemOpen
  \bibfield  {author} {\bibinfo {author} {\bibfnamefont {B.}~\bibnamefont
  {Miksch}}, \bibinfo {author} {\bibfnamefont {M.}~\bibnamefont {Dressel}},\
  and\ \bibinfo {author} {\bibfnamefont {M.}~\bibnamefont {Scheffler}},\
  }\bibfield  {title} {\bibinfo {title} {{Cryogenic frequency-domain electron
  spin resonance spectrometer based on coplanar waveguides and field
  modulation}},\ }\href {https://doi.org/10.1063/1.5141461} {\bibfield
  {journal} {\bibinfo  {journal} {Review of Scientific Instruments}\ }\textbf
  {\bibinfo {volume} {91}},\ \bibinfo {pages} {025106} (\bibinfo {year}
  {2020})}\BibitemShut {NoStop}%
\bibitem [{\citenamefont {Maier-Flaig}\ \emph {et~al.}(2017)\citenamefont
  {Maier-Flaig}, \citenamefont {Klingler}, \citenamefont {Dubs}, \citenamefont
  {Surzhenko}, \citenamefont {Gross}, \citenamefont {Weiler}, \citenamefont
  {Huebl},\ and\ \citenamefont {Goennenwein}}]{MaierFlaig2017}%
  \BibitemOpen
  \bibfield  {author} {\bibinfo {author} {\bibfnamefont {H.}~\bibnamefont
  {Maier-Flaig}}, \bibinfo {author} {\bibfnamefont {S.}~\bibnamefont
  {Klingler}}, \bibinfo {author} {\bibfnamefont {C.}~\bibnamefont {Dubs}},
  \bibinfo {author} {\bibfnamefont {O.}~\bibnamefont {Surzhenko}}, \bibinfo
  {author} {\bibfnamefont {R.}~\bibnamefont {Gross}}, \bibinfo {author}
  {\bibfnamefont {M.}~\bibnamefont {Weiler}}, \bibinfo {author} {\bibfnamefont
  {H.}~\bibnamefont {Huebl}},\ and\ \bibinfo {author} {\bibfnamefont
  {S.~T.~B.}\ \bibnamefont {Goennenwein}},\ }\bibfield  {title} {\bibinfo
  {title} {Temperature-dependent magnetic damping of yttrium iron garnet
  spheres},\ }\href {https://doi.org/10.1103/PhysRevB.95.214423} {\bibfield
  {journal} {\bibinfo  {journal} {Phys. Rev. B}\ }\textbf {\bibinfo {volume}
  {95}},\ \bibinfo {pages} {214423} (\bibinfo {year} {2017})}\BibitemShut
  {NoStop}%
\bibitem [{\citenamefont {Klingler}\ \emph {et~al.}(2017)\citenamefont
  {Klingler}, \citenamefont {Maier-Flaig}, \citenamefont {Dubs}, \citenamefont
  {Surzhenko}, \citenamefont {Gross}, \citenamefont {Huebl}, \citenamefont
  {Goennenwein},\ and\ \citenamefont {Weiler}}]{Klingler2017}%
  \BibitemOpen
  \bibfield  {author} {\bibinfo {author} {\bibfnamefont {S.}~\bibnamefont
  {Klingler}}, \bibinfo {author} {\bibfnamefont {H.}~\bibnamefont
  {Maier-Flaig}}, \bibinfo {author} {\bibfnamefont {C.}~\bibnamefont {Dubs}},
  \bibinfo {author} {\bibfnamefont {O.}~\bibnamefont {Surzhenko}}, \bibinfo
  {author} {\bibfnamefont {R.}~\bibnamefont {Gross}}, \bibinfo {author}
  {\bibfnamefont {H.}~\bibnamefont {Huebl}}, \bibinfo {author} {\bibfnamefont
  {S.~T.~B.}\ \bibnamefont {Goennenwein}},\ and\ \bibinfo {author}
  {\bibfnamefont {M.}~\bibnamefont {Weiler}},\ }\bibfield  {title} {\bibinfo
  {title} {Gilbert damping of magnetostatic modes in a yttrium iron garnet
  sphere},\ }\href {https://doi.org/10.1063/1.4977423} {\bibfield  {journal}
  {\bibinfo  {journal} {Applied Physics Letters}\ }\textbf {\bibinfo {volume}
  {110}},\ \bibinfo {pages} {092409} (\bibinfo {year} {2017})}\BibitemShut
  {NoStop}%
\bibitem [{\citenamefont {Kukharchyk}\ \emph {et~al.}(2018)\citenamefont
  {Kukharchyk}, \citenamefont {Sholokhov}, \citenamefont {Morozov},
  \citenamefont {Korableva}, \citenamefont {Kalachev},\ and\ \citenamefont
  {Bushev}}]{Kukharchyk2018}%
  \BibitemOpen
  \bibfield  {author} {\bibinfo {author} {\bibfnamefont {N.}~\bibnamefont
  {Kukharchyk}}, \bibinfo {author} {\bibfnamefont {D.}~\bibnamefont
  {Sholokhov}}, \bibinfo {author} {\bibfnamefont {O.}~\bibnamefont {Morozov}},
  \bibinfo {author} {\bibfnamefont {S.~L.}\ \bibnamefont {Korableva}}, \bibinfo
  {author} {\bibfnamefont {A.~A.}\ \bibnamefont {Kalachev}},\ and\ \bibinfo
  {author} {\bibfnamefont {P.~A.}\ \bibnamefont {Bushev}},\ }\bibfield  {title}
  {\bibinfo {title} {{Optical coherence of
  $^{166}\mathrm{Er}\!:^{7}\!\mathrm{LiYF}_{4}$ crystal below 1 K}},\ }\href
  {https://doi.org/10.1088/1367-2630/aaa7e4} {\bibfield  {journal} {\bibinfo
  {journal} {New Journal of Physics}\ }\textbf {\bibinfo {volume} {20}},\
  \bibinfo {pages} {023044} (\bibinfo {year} {2018})}\BibitemShut {NoStop}%
\bibitem [{\citenamefont {Vogl}(2023)}]{Vogl2023}%
  \BibitemOpen
  \bibfield  {author} {\bibinfo {author} {\bibfnamefont {L.}~\bibnamefont
  {Vogl}},\ }\href@noop {} {\bibinfo {title} {Generating small magnetic fields
  inside an open-end magnetic shielding with a superconducting solenoid
  magnet}} (\bibinfo {year} {2023})\BibitemShut {NoStop}%
\bibitem [{\citenamefont {Nehrkorn}\ \emph {et~al.}(2015)\citenamefont
  {Nehrkorn}, \citenamefont {Schnegg}, \citenamefont {Holldack},\ and\
  \citenamefont {Stoll}}]{Nehrkorn2015}%
  \BibitemOpen
  \bibfield  {author} {\bibinfo {author} {\bibfnamefont {J.}~\bibnamefont
  {Nehrkorn}}, \bibinfo {author} {\bibfnamefont {A.}~\bibnamefont {Schnegg}},
  \bibinfo {author} {\bibfnamefont {K.}~\bibnamefont {Holldack}},\ and\
  \bibinfo {author} {\bibfnamefont {S.}~\bibnamefont {Stoll}},\ }\bibfield
  {title} {\bibinfo {title} {{General Magnetic Transition Dipole Moments for
  Electron Paramagnetic Resonance}},\ }\href
  {https://doi.org/10.1103/PhysRevLett.114.010801} {\bibfield  {journal}
  {\bibinfo  {journal} {Phys. Rev. Lett.}\ }\textbf {\bibinfo {volume} {114}},\
  \bibinfo {pages} {010801} (\bibinfo {year} {2015})}\BibitemShut {NoStop}%
\bibitem [{\citenamefont {Momma}\ and\ \citenamefont
  {Izumi}(2011)}]{Momma2011}%
  \BibitemOpen
  \bibfield  {author} {\bibinfo {author} {\bibfnamefont {K.}~\bibnamefont
  {Momma}}\ and\ \bibinfo {author} {\bibfnamefont {F.}~\bibnamefont {Izumi}},\
  }\bibfield  {title} {\bibinfo {title} {{{\it VESTA3} for three-dimensional
  visualization of crystal, volumetric and morphology data}},\ }\href
  {https://doi.org/10.1107/S0021889811038970} {\bibfield  {journal} {\bibinfo
  {journal} {Journal of Applied Crystallography}\ }\textbf {\bibinfo {volume}
  {44}},\ \bibinfo {pages} {1272} (\bibinfo {year} {2011})}\BibitemShut
  {NoStop}%
\bibitem [{\citenamefont {Goryunov}\ \emph {et~al.}(1992)\citenamefont
  {Goryunov}, \citenamefont {Popov}, \citenamefont {Khajdukov},\ and\
  \citenamefont {Fedorov}}]{Goryunov1992}%
  \BibitemOpen
  \bibfield  {author} {\bibinfo {author} {\bibfnamefont {A.}~\bibnamefont
  {Goryunov}}, \bibinfo {author} {\bibfnamefont {A.}~\bibnamefont {Popov}},
  \bibinfo {author} {\bibfnamefont {N.}~\bibnamefont {Khajdukov}},\ and\
  \bibinfo {author} {\bibfnamefont {P.}~\bibnamefont {Fedorov}},\ }\bibfield
  {title} {\bibinfo {title} {{Crystal structure of lithium and yttrium complex
  fluorides}},\ }\href
  {https://doi.org/https://doi.org/10.1016/0025-5408(92)90215-L} {\bibfield
  {journal} {\bibinfo  {journal} {Materials Research Bulletin}\ }\textbf
  {\bibinfo {volume} {27}},\ \bibinfo {pages} {213–220} (\bibinfo {year}
  {1992})}\BibitemShut {NoStop}%
\bibitem [{\citenamefont {Sattler}\ and\ \citenamefont
  {Nemarich}(1971)}]{Sattler1971}%
  \BibitemOpen
  \bibfield  {author} {\bibinfo {author} {\bibfnamefont {J.~P.}\ \bibnamefont
  {Sattler}}\ and\ \bibinfo {author} {\bibfnamefont {J.}~\bibnamefont
  {Nemarich}},\ }\bibfield  {title} {\bibinfo {title}
  {{Electron-Paramagnetic-Resonance Spectra of $\mathrm{Nd}^{3+}$,
  $\mathrm{Dy}^{3+}$, $\mathrm{Er}^{3+}$, and $\mathrm{Yb}^{3+}$ in Lithium
  Yttrium Fluoride}},\ }\href {https://doi.org/10.1103/PhysRevB.4.1} {\bibfield
   {journal} {\bibinfo  {journal} {Phys. Rev. B}\ }\textbf {\bibinfo {volume}
  {4}},\ \bibinfo {pages} {1–5} (\bibinfo {year} {1971})}\BibitemShut
  {NoStop}%
\bibitem [{\citenamefont {Chukalina}\ and\ \citenamefont
  {Popova}(1999)}]{Chukalina1999}%
  \BibitemOpen
  \bibfield  {author} {\bibinfo {author} {\bibfnamefont {E.}~\bibnamefont
  {Chukalina}}\ and\ \bibinfo {author} {\bibfnamefont {M.}~\bibnamefont
  {Popova}},\ }\bibfield  {title} {\bibinfo {title} {{Hyperfine structure of
  infrared transitions in $\mathrm{LiYF}_{4}\!:\!\mathrm{Er}^{3+}$}},\ }\href
  {https://doi.org/https://doi.org/10.1016/S0375-9601(99)00687-8} {\bibfield
  {journal} {\bibinfo  {journal} {Physics Letters A}\ }\textbf {\bibinfo
  {volume} {262}},\ \bibinfo {pages} {191–194} (\bibinfo {year}
  {1999})}\BibitemShut {NoStop}%
\bibitem [{\citenamefont {Vishwamittar}\ and\ \citenamefont
  {Puri}(1974)}]{Vishwamittar1974}%
  \BibitemOpen
  \bibfield  {author} {\bibinfo {author} {\bibnamefont {Vishwamittar}}\ and\
  \bibinfo {author} {\bibfnamefont {S.~P.}\ \bibnamefont {Puri}},\ }\bibfield
  {title} {\bibinfo {title} {{Interpretation of the crystal-field parameters in
  a rare-earth substituted $\mathrm{LiYF}_{4}$ crystal}},\ }\href
  {https://doi.org/10.1088/0022-3719/7/7/025} {\bibfield  {journal} {\bibinfo
  {journal} {Journal of Physics C: Solid State Physics}\ }\textbf {\bibinfo
  {volume} {7}},\ \bibinfo {pages} {1337} (\bibinfo {year} {1974})}\BibitemShut
  {NoStop}%
\bibitem [{\citenamefont {Macfarlane}\ \emph {et~al.}(1992)\citenamefont
  {Macfarlane}, \citenamefont {Cassanho},\ and\ \citenamefont
  {Meltzer}}]{Macfarlane1992}%
  \BibitemOpen
  \bibfield  {author} {\bibinfo {author} {\bibfnamefont {R.~M.}\ \bibnamefont
  {Macfarlane}}, \bibinfo {author} {\bibfnamefont {A.}~\bibnamefont
  {Cassanho}},\ and\ \bibinfo {author} {\bibfnamefont {R.~S.}\ \bibnamefont
  {Meltzer}},\ }\bibfield  {title} {\bibinfo {title} {{Inhomogeneous broadening
  by nuclear spin fields: A new limit for optical transitions in solids}},\
  }\href {https://doi.org/10.1103/PhysRevLett.69.542} {\bibfield  {journal}
  {\bibinfo  {journal} {Phys. Rev. Lett.}\ }\textbf {\bibinfo {volume} {69}},\
  \bibinfo {pages} {542–545} (\bibinfo {year} {1992})}\BibitemShut {NoStop}%
\bibitem [{\citenamefont {Abragam}\ and\ \citenamefont
  {Bleaney}(2012)}]{AbragamBleaney}%
  \BibitemOpen
  \bibfield  {author} {\bibinfo {author} {\bibfnamefont {A.}~\bibnamefont
  {Abragam}}\ and\ \bibinfo {author} {\bibfnamefont {B.}~\bibnamefont
  {Bleaney}},\ }\href@noop {} {\emph {\bibinfo {title} {{Electron Paramagnetic
  Resonance of Transition Ions}}}},\ Oxford Classic Texts in the Physical
  Sciences\ (\bibinfo  {publisher} {OUP Oxford},\ \bibinfo {year}
  {2012})\BibitemShut {NoStop}%
\bibitem [{\citenamefont {Stoll}\ and\ \citenamefont
  {Goldfarb}(2017)}]{Stoll2017}%
  \BibitemOpen
  \bibfield  {author} {\bibinfo {author} {\bibfnamefont {S.}~\bibnamefont
  {Stoll}}\ and\ \bibinfo {author} {\bibfnamefont {D.}~\bibnamefont
  {Goldfarb}},\ }\bibinfo {title} {{EPR Interactions – Nuclear Quadrupole
  Couplings}},\ in\ \href
  {https://doi.org/https://doi.org/10.1002/9780470034590.emrstm1504} {\emph
  {\bibinfo {booktitle} {eMagRes}}}\ (\bibinfo  {publisher} {John Wiley \&
  Sons, Ltd},\ \bibinfo {year} {2017})\ pp.\ \bibinfo {pages}
  {495--510}\BibitemShut {NoStop}%
\bibitem [{\citenamefont {Stoll}(2014)}]{Stoll2014}%
  \BibitemOpen
  \bibfield  {author} {\bibinfo {author} {\bibfnamefont {S.}~\bibnamefont
  {Stoll}},\ }\bibinfo {title} {Computational modeling and least-squares
  fitting of $\mathrm{EPR}$ spectra},\ in\ \href
  {https://doi.org/https://doi.org/10.1002/9783527672431.ch3} {\emph {\bibinfo
  {booktitle} {Multifrequency Electron Paramagnetic Resonance}}}\ (\bibinfo
  {publisher} {John Wiley \& Sons, Ltd},\ \bibinfo {year} {2014})\
  Chap.~\bibinfo {chapter} {3}, pp.\ \bibinfo {pages} {69--138}\BibitemShut
  {NoStop}%
\bibitem [{\citenamefont {Brown}\ \emph {et~al.}(1969)\citenamefont {Brown},
  \citenamefont {Roots},\ and\ \citenamefont {Shand}}]{Brown1969}%
  \BibitemOpen
  \bibfield  {author} {\bibinfo {author} {\bibfnamefont {M.~R.}\ \bibnamefont
  {Brown}}, \bibinfo {author} {\bibfnamefont {K.~G.}\ \bibnamefont {Roots}},\
  and\ \bibinfo {author} {\bibfnamefont {W.~A.}\ \bibnamefont {Shand}},\
  }\bibfield  {title} {\bibinfo {title} {{Energy levels of $\mathrm{Er}^{3+}$
  in $\mathrm{LiYF}_{4}$}},\ }\href {https://doi.org/10.1088/0022-3719/2/4/304}
  {\bibfield  {journal} {\bibinfo  {journal} {Journal of Physics C: Solid State
  Physics}\ }\textbf {\bibinfo {volume} {2}},\ \bibinfo {pages} {593} (\bibinfo
  {year} {1969})}\BibitemShut {NoStop}%
\bibitem [{\citenamefont {Guedes}\ \emph {et~al.}(2002)\citenamefont {Guedes},
  \citenamefont {Krambrock},\ and\ \citenamefont {Gesland}}]{Guedes2002}%
  \BibitemOpen
  \bibfield  {author} {\bibinfo {author} {\bibfnamefont {K.}~\bibnamefont
  {Guedes}}, \bibinfo {author} {\bibfnamefont {K.}~\bibnamefont {Krambrock}},\
  and\ \bibinfo {author} {\bibfnamefont {J.}~\bibnamefont {Gesland}},\
  }\bibfield  {title} {\bibinfo {title} {{Identification of trivalent rare
  earth impurities in $\mathrm{YF}_{3}$ , $\mathrm{LuF}_{3}$ and
  $\mathrm{LiYF}_{4}$ by electron paramagnetic resonance}},\ }\href
  {https://doi.org/10.1016/S0925-8388(02)00362-6} {\bibfield  {journal}
  {\bibinfo  {journal} {J. Alloys Compd.}\ }\textbf {\bibinfo {volume} {344}},\
  \bibinfo {pages} {251–254} (\bibinfo {year} {2002})}\BibitemShut {NoStop}%
\bibitem [{\citenamefont {Marino}\ \emph {et~al.}(2016)\citenamefont {Marino},
  \citenamefont {Lorgeré}, \citenamefont {Guillot-Noël}, \citenamefont
  {Vezin}, \citenamefont {Toncelli}, \citenamefont {Tonelli}, \citenamefont
  {{Le Gouët}},\ and\ \citenamefont {Goldner}}]{Marino2016}%
  \BibitemOpen
  \bibfield  {author} {\bibinfo {author} {\bibfnamefont {R.}~\bibnamefont
  {Marino}}, \bibinfo {author} {\bibfnamefont {I.}~\bibnamefont {Lorgeré}},
  \bibinfo {author} {\bibfnamefont {O.}~\bibnamefont {Guillot-Noël}}, \bibinfo
  {author} {\bibfnamefont {H.}~\bibnamefont {Vezin}}, \bibinfo {author}
  {\bibfnamefont {A.}~\bibnamefont {Toncelli}}, \bibinfo {author}
  {\bibfnamefont {M.}~\bibnamefont {Tonelli}}, \bibinfo {author} {\bibfnamefont
  {J.-L.}\ \bibnamefont {{Le Gouët}}},\ and\ \bibinfo {author} {\bibfnamefont
  {P.}~\bibnamefont {Goldner}},\ }\bibfield  {title} {\bibinfo {title} {{Energy
  level structure and optical dephasing under magnetic field in
  $\mathrm{Er}^{3+}$\!:$\mathrm{LiYF}_{4}$ at 1.5 $\mu$m}},\ }\href
  {https://doi.org/https://doi.org/10.1016/j.jlumin.2015.03.003} {\bibfield
  {journal} {\bibinfo  {journal} {Journal of Luminescence}\ }\textbf {\bibinfo
  {volume} {169}},\ \bibinfo {pages} {478–482} (\bibinfo {year} {2016})},\
  \bibinfo {note} {the 17th International Conference on Luminescence and
  Optical Spectroscopy of Condensed Matter (ICL'14)}\BibitemShut {NoStop}%
\bibitem [{\citenamefont {Lisin}\ \emph {et~al.}(2019)\citenamefont {Lisin},
  \citenamefont {Shegeda}, \citenamefont {Samartsev},\ and\ \citenamefont
  {Korableva}}]{Lisin2019}%
  \BibitemOpen
  \bibfield  {author} {\bibinfo {author} {\bibfnamefont {V.~N.}\ \bibnamefont
  {Lisin}}, \bibinfo {author} {\bibfnamefont {A.~M.}\ \bibnamefont {Shegeda}},
  \bibinfo {author} {\bibfnamefont {V.~V.}\ \bibnamefont {Samartsev}},\ and\
  \bibinfo {author} {\bibfnamefont {S.~L.}\ \bibnamefont {Korableva}},\
  }\bibfield  {title} {\bibinfo {title} {{Incoherent photon echo in
  $^{7}\mathrm{LiYF}_{4}\!:\!^{167}\mathrm{Er}^{3+}$}},\ }\href
  {https://doi.org/10.1088/1555-6611/ab4bd4} {\bibfield  {journal} {\bibinfo
  {journal} {Laser Physics}\ }\textbf {\bibinfo {volume} {29}},\ \bibinfo
  {pages} {124005} (\bibinfo {year} {2019})}\BibitemShut {NoStop}%
\bibitem [{\citenamefont {Gerasimov}\ \emph {et~al.}(2016)\citenamefont
  {Gerasimov}, \citenamefont {Minnegaliev}, \citenamefont {Malkin},
  \citenamefont {Baibekov},\ and\ \citenamefont {Moiseev}}]{Gerasimov2016}%
  \BibitemOpen
  \bibfield  {author} {\bibinfo {author} {\bibfnamefont {K.~I.}\ \bibnamefont
  {Gerasimov}}, \bibinfo {author} {\bibfnamefont {M.~M.}\ \bibnamefont
  {Minnegaliev}}, \bibinfo {author} {\bibfnamefont {B.~Z.}\ \bibnamefont
  {Malkin}}, \bibinfo {author} {\bibfnamefont {E.~I.}\ \bibnamefont
  {Baibekov}},\ and\ \bibinfo {author} {\bibfnamefont {S.~A.}\ \bibnamefont
  {Moiseev}},\ }\bibfield  {title} {\bibinfo {title} {{High-resolution
  magneto-optical spectroscopy of
  $^{7}\mathrm{LiYF}_{4}\!:\!^{167}\!\mathrm{Er}^{3+},\!^{166}\mathrm{Er}^{3+}$
  and analysis of hyperfine structure of ultranarrow optical transitions}},\
  }\href {https://doi.org/10.1103/PhysRevB.94.054429} {\bibfield  {journal}
  {\bibinfo  {journal} {Phys. Rev. B}\ }\textbf {\bibinfo {volume} {94}},\
  \bibinfo {pages} {054429} (\bibinfo {year} {2016})}\BibitemShut {NoStop}%
\bibitem [{\citenamefont {Kukharchyk}\ \emph {et~al.}(2020)\citenamefont
  {Kukharchyk}, \citenamefont {Sholokhov}, \citenamefont {Morozov},
  \citenamefont {Korableva}, \citenamefont {Kalachev},\ and\ \citenamefont
  {Bushev}}]{Kukharchyk2020}%
  \BibitemOpen
  \bibfield  {author} {\bibinfo {author} {\bibfnamefont {N.}~\bibnamefont
  {Kukharchyk}}, \bibinfo {author} {\bibfnamefont {D.}~\bibnamefont
  {Sholokhov}}, \bibinfo {author} {\bibfnamefont {O.}~\bibnamefont {Morozov}},
  \bibinfo {author} {\bibfnamefont {S.~L.}\ \bibnamefont {Korableva}}, \bibinfo
  {author} {\bibfnamefont {A.~A.}\ \bibnamefont {Kalachev}},\ and\ \bibinfo
  {author} {\bibfnamefont {P.~A.}\ \bibnamefont {Bushev}},\ }\bibfield  {title}
  {\bibinfo {title} {{Electromagnetically induced transparency in a
  mono-isotopic $\mathrm{^{167}Er\!:\!^{7}LiYF_{4}}$ crystal below 1 Kelvin:
  microwave photonics approach}},\ }\href {https://doi.org/10.1364/OE.400222}
  {\bibfield  {journal} {\bibinfo  {journal} {Opt. Express}\ }\textbf {\bibinfo
  {volume} {28}},\ \bibinfo {pages} {29166–29177} (\bibinfo {year}
  {2020})}\BibitemShut {NoStop}%
\bibitem [{\citenamefont {Popova}\ \emph {et~al.}(2000)\citenamefont {Popova},
  \citenamefont {Chukalina}, \citenamefont {Malkin},\ and\ \citenamefont
  {Saikin}}]{Popova2000}%
  \BibitemOpen
  \bibfield  {author} {\bibinfo {author} {\bibfnamefont {M.~N.}\ \bibnamefont
  {Popova}}, \bibinfo {author} {\bibfnamefont {E.~P.}\ \bibnamefont
  {Chukalina}}, \bibinfo {author} {\bibfnamefont {B.~Z.}\ \bibnamefont
  {Malkin}},\ and\ \bibinfo {author} {\bibfnamefont {S.~K.}\ \bibnamefont
  {Saikin}},\ }\bibfield  {title} {\bibinfo {title} {{Experimental and
  theoretical study of the crystal-field levels and hyperfine and
  electron-phonon interactions in $\mathrm{LiYF}_{4}\!:\!\mathrm{Er}^{3+}$}},\
  }\href {https://doi.org/10.1103/PhysRevB.61.7421} {\bibfield  {journal}
  {\bibinfo  {journal} {Phys. Rev. B}\ }\textbf {\bibinfo {volume} {61}},\
  \bibinfo {pages} {7421–7427} (\bibinfo {year} {2000})}\BibitemShut
  {NoStop}%
\bibitem [{\citenamefont {Wu}(2004)}]{Wu2004}%
  \BibitemOpen
  \bibfield  {author} {\bibinfo {author} {\bibfnamefont {S.-Y.}\ \bibnamefont
  {Wu}},\ }\bibfield  {title} {\bibinfo {title} {{Theoretical investigations on
  the spin Hamiltonian parameters for $\mathrm{Er}^{3+}$ ion in
  $\mathrm{LiYF}_{4}$}},\ }\href
  {https://doi.org/https://doi.org/10.1016/j.optmat.2004.02.012} {\bibfield
  {journal} {\bibinfo  {journal} {Optical Materials}\ }\textbf {\bibinfo
  {volume} {27}},\ \bibinfo {pages} {99} (\bibinfo {year} {2004})}\BibitemShut
  {NoStop}%
\bibitem [{\citenamefont {Elliott}\ \emph {et~al.}(1953)\citenamefont
  {Elliott}, \citenamefont {Stevens},\ and\ \citenamefont
  {Pryce}}]{Elliott1953}%
  \BibitemOpen
  \bibfield  {author} {\bibinfo {author} {\bibfnamefont {R.~J.}\ \bibnamefont
  {Elliott}}, \bibinfo {author} {\bibfnamefont {K.~W.~H.}\ \bibnamefont
  {Stevens}},\ and\ \bibinfo {author} {\bibfnamefont {M.~H.~L.}\ \bibnamefont
  {Pryce}},\ }\bibfield  {title} {\bibinfo {title} {{The theory of magnetic
  resonance experiments on salts of the rare earths}},\ }\href
  {https://doi.org/10.1098/rspa.1953.0124} {\bibfield  {journal} {\bibinfo
  {journal} {Proc. R. Soc. Lond. A}\ }\textbf {\bibinfo {volume} {218}},\
  \bibinfo {pages} {553–566} (\bibinfo {year} {1953})}\BibitemShut {NoStop}%
\bibitem [{\citenamefont {Abraham}\ \emph {et~al.}(1983)\citenamefont
  {Abraham}, \citenamefont {Boatner}, \citenamefont {Ramey},\ and\
  \citenamefont {Rappaz}}]{Abraham1983}%
  \BibitemOpen
  \bibfield  {author} {\bibinfo {author} {\bibfnamefont {M.~M.}\ \bibnamefont
  {Abraham}}, \bibinfo {author} {\bibfnamefont {L.~A.}\ \bibnamefont
  {Boatner}}, \bibinfo {author} {\bibfnamefont {J.~O.}\ \bibnamefont {Ramey}},\
  and\ \bibinfo {author} {\bibfnamefont {M.}~\bibnamefont {Rappaz}},\
  }\bibfield  {title} {\bibinfo {title} {{An EPR study of rare‐earth
  impurities in single crystals of the zircon‐structure orthophosphates
  $\mathrm{ScPO}_{4}$, $\mathrm{YPO}_{4}$, and $\mathrm{LuPO}_{4}$}},\ }\href
  {https://doi.org/10.1063/1.444516} {\bibfield  {journal} {\bibinfo  {journal}
  {The Journal of Chemical Physics}\ }\textbf {\bibinfo {volume} {78}},\
  \bibinfo {pages} {3–10} (\bibinfo {year} {1983})}\BibitemShut {NoStop}%
\bibitem [{\citenamefont {Macfarlane}\ \emph {et~al.}(1998)\citenamefont
  {Macfarlane}, \citenamefont {Meltzer},\ and\ \citenamefont
  {Malkin}}]{Macfarlane1998}%
  \BibitemOpen
  \bibfield  {author} {\bibinfo {author} {\bibfnamefont {R.~M.}\ \bibnamefont
  {Macfarlane}}, \bibinfo {author} {\bibfnamefont {R.~S.}\ \bibnamefont
  {Meltzer}},\ and\ \bibinfo {author} {\bibfnamefont {B.~Z.}\ \bibnamefont
  {Malkin}},\ }\bibfield  {title} {\bibinfo {title} {Optical measurement of the
  isotope shifts and hyperfine and superhyperfine interactions of nd in the
  solid state},\ }\href {https://doi.org/10.1103/PhysRevB.58.5692} {\bibfield
  {journal} {\bibinfo  {journal} {Phys. Rev. B}\ }\textbf {\bibinfo {volume}
  {58}},\ \bibinfo {pages} {5692–5700} (\bibinfo {year} {1998})}\BibitemShut
  {NoStop}%
\bibitem [{\citenamefont {Kukharchyk}\ \emph {et~al.}(2021)\citenamefont
  {Kukharchyk}, \citenamefont {Sholokhov}, \citenamefont {Kalachev},\ and\
  \citenamefont {Bushev}}]{Kukharchyk2021}%
  \BibitemOpen
  \bibfield  {author} {\bibinfo {author} {\bibfnamefont {N.}~\bibnamefont
  {Kukharchyk}}, \bibinfo {author} {\bibfnamefont {D.}~\bibnamefont
  {Sholokhov}}, \bibinfo {author} {\bibfnamefont {A.~A.}\ \bibnamefont
  {Kalachev}},\ and\ \bibinfo {author} {\bibfnamefont {P.~A.}\ \bibnamefont
  {Bushev}},\ }\href {https://arxiv.org/abs/1910.03096} {\bibinfo {title}
  {Enhancement of optical coherence in
  $^{167}\mathrm{Er}\!:\!\mathrm{Y}_{2}\mathrm{SiO}_{5}$ crystal at sub-kelvin
  temperatures}} (\bibinfo {year} {2021}),\ \Eprint
  {https://arxiv.org/abs/1910.03096} {arXiv:1910.03096 [quant-ph]} \BibitemShut
  {NoStop}%
\bibitem [{\citenamefont {Mitrikas}\ \emph {et~al.}(2020)\citenamefont
  {Mitrikas}, \citenamefont {Sanakis},\ and\ \citenamefont
  {Ioannidis}}]{Mitrikas2020}%
  \BibitemOpen
  \bibfield  {author} {\bibinfo {author} {\bibfnamefont {G.}~\bibnamefont
  {Mitrikas}}, \bibinfo {author} {\bibfnamefont {Y.}~\bibnamefont {Sanakis}},\
  and\ \bibinfo {author} {\bibfnamefont {N.}~\bibnamefont {Ioannidis}},\
  }\bibfield  {title} {\bibinfo {title} {{Parallel-Mode EPR of Atomic Hydrogen
  Encapsulated in POSS Cages}},\ }\href
  {https://doi.org/10.1007/s00723-020-01263-5} {\bibfield  {journal} {\bibinfo
  {journal} {Applied Magnetic Resonance}\ }\textbf {\bibinfo {volume} {51}},\
  \bibinfo {pages} {1451–1466} (\bibinfo {year} {2020})}\BibitemShut
  {NoStop}%
\bibitem [{\citenamefont {Rudowicz}\ \emph {et~al.}(2021)\citenamefont
  {Rudowicz}, \citenamefont {Cecot},\ and\ \citenamefont
  {Krasowski}}]{Rudowicz2021}%
  \BibitemOpen
  \bibfield  {author} {\bibinfo {author} {\bibfnamefont {C.}~\bibnamefont
  {Rudowicz}}, \bibinfo {author} {\bibfnamefont {P.}~\bibnamefont {Cecot}},\
  and\ \bibinfo {author} {\bibfnamefont {M.}~\bibnamefont {Krasowski}},\
  }\bibfield  {title} {\bibinfo {title} {{Selection rules in electron magnetic
  resonance (EMR) spectroscopy and related techniques: Fundamentals and
  applications to modern case systems}},\ }\href
  {https://doi.org/https://doi.org/10.1016/j.physb.2021.412863} {\bibfield
  {journal} {\bibinfo  {journal} {Physica B: Condensed Matter}\ }\textbf
  {\bibinfo {volume} {608}},\ \bibinfo {pages} {412863} (\bibinfo {year}
  {2021})}\BibitemShut {NoStop}%
\bibitem [{\citenamefont {Mims}(1965)}]{Mims1965}%
  \BibitemOpen
  \bibfield  {author} {\bibinfo {author} {\bibfnamefont {W.~B.}\ \bibnamefont
  {Mims}},\ }\bibfield  {title} {\bibinfo {title} {{Electric Field Shift in
  Paramagnetic Resonance for Four Ions in a Calcium Tungstate Lattice}},\
  }\href {https://doi.org/10.1103/PhysRev.140.A531} {\bibfield  {journal}
  {\bibinfo  {journal} {Phys. Rev.}\ }\textbf {\bibinfo {volume} {140}},\
  \bibinfo {pages} {A531} (\bibinfo {year} {1965})}\BibitemShut {NoStop}%
\bibitem [{\citenamefont {Fraval}\ \emph {et~al.}(2004)\citenamefont {Fraval},
  \citenamefont {Sellars},\ and\ \citenamefont {Longdell}}]{Fraval2004}%
  \BibitemOpen
  \bibfield  {author} {\bibinfo {author} {\bibfnamefont {E.}~\bibnamefont
  {Fraval}}, \bibinfo {author} {\bibfnamefont {M.~J.}\ \bibnamefont
  {Sellars}},\ and\ \bibinfo {author} {\bibfnamefont {J.~J.}\ \bibnamefont
  {Longdell}},\ }\bibfield  {title} {\bibinfo {title} {Method of extending
  hyperfine coherence times in
  $\mathrm{Pr}^{3+}$\!:$\mathrm{Y}_{2}\mathrm{SiO}_{5}$},\ }\href
  {https://doi.org/10.1103/PhysRevLett.92.077601} {\bibfield  {journal}
  {\bibinfo  {journal} {Phys. Rev. Lett.}\ }\textbf {\bibinfo {volume} {92}},\
  \bibinfo {pages} {077601} (\bibinfo {year} {2004})}\BibitemShut {NoStop}%
\bibitem [{\citenamefont {Fraval}\ \emph {et~al.}(2005)\citenamefont {Fraval},
  \citenamefont {Sellars},\ and\ \citenamefont {Longdell}}]{Fraval2005}%
  \BibitemOpen
  \bibfield  {author} {\bibinfo {author} {\bibfnamefont {E.}~\bibnamefont
  {Fraval}}, \bibinfo {author} {\bibfnamefont {M.~J.}\ \bibnamefont
  {Sellars}},\ and\ \bibinfo {author} {\bibfnamefont {J.~J.}\ \bibnamefont
  {Longdell}},\ }\bibfield  {title} {\bibinfo {title} {{Dynamic Decoherence
  Control of a Solid-State Nuclear-Quadrupole Qubit}},\ }\href
  {https://doi.org/10.1103/PhysRevLett.95.030506} {\bibfield  {journal}
  {\bibinfo  {journal} {Phys. Rev. Lett.}\ }\textbf {\bibinfo {volume} {95}},\
  \bibinfo {pages} {030506} (\bibinfo {year} {2005})}\BibitemShut {NoStop}%
\bibitem [{\citenamefont {Bilzer}\ \emph {et~al.}(2007)\citenamefont {Bilzer},
  \citenamefont {Devolder}, \citenamefont {Crozat}, \citenamefont {Chappert},
  \citenamefont {Cardoso},\ and\ \citenamefont {Freitas}}]{Bilzer2007}%
  \BibitemOpen
  \bibfield  {author} {\bibinfo {author} {\bibfnamefont {C.}~\bibnamefont
  {Bilzer}}, \bibinfo {author} {\bibfnamefont {T.}~\bibnamefont {Devolder}},
  \bibinfo {author} {\bibfnamefont {P.}~\bibnamefont {Crozat}}, \bibinfo
  {author} {\bibfnamefont {C.}~\bibnamefont {Chappert}}, \bibinfo {author}
  {\bibfnamefont {S.}~\bibnamefont {Cardoso}},\ and\ \bibinfo {author}
  {\bibfnamefont {P.~P.}\ \bibnamefont {Freitas}},\ }\bibfield  {title}
  {\bibinfo {title} {{Vector network analyzer ferromagnetic resonance of thin
  films on coplanar waveguides: Comparison of different evaluation methods}},\
  }\href {https://doi.org/10.1063/1.2716995} {\bibfield  {journal} {\bibinfo
  {journal} {Journal of Applied Physics}\ }\textbf {\bibinfo {volume} {101}},\
  \bibinfo {pages} {074505} (\bibinfo {year} {2007})}\BibitemShut {NoStop}%
\bibitem [{\citenamefont {Velluire-Pellat}\ \emph {et~al.}(2023)\citenamefont
  {Velluire-Pellat}, \citenamefont {Maréchal}, \citenamefont {Moulonguet},
  \citenamefont {Saïz}, \citenamefont {Ménard}, \citenamefont {Kozlov},
  \citenamefont {Couëdo}, \citenamefont {Amari}, \citenamefont {Medous},
  \citenamefont {Paris}, \citenamefont {Hostein}, \citenamefont {Lesueur},
  \citenamefont {Feuillet-Palma},\ and\ \citenamefont
  {Bergeal}}]{VelluirePellat2023}%
  \BibitemOpen
  \bibfield  {author} {\bibinfo {author} {\bibfnamefont {Z.}~\bibnamefont
  {Velluire-Pellat}}, \bibinfo {author} {\bibfnamefont {E.}~\bibnamefont
  {Maréchal}}, \bibinfo {author} {\bibfnamefont {N.}~\bibnamefont
  {Moulonguet}}, \bibinfo {author} {\bibfnamefont {G.}~\bibnamefont {Saïz}},
  \bibinfo {author} {\bibfnamefont {G.~C.}\ \bibnamefont {Ménard}}, \bibinfo
  {author} {\bibfnamefont {S.}~\bibnamefont {Kozlov}}, \bibinfo {author}
  {\bibfnamefont {F.}~\bibnamefont {Couëdo}}, \bibinfo {author} {\bibfnamefont
  {P.}~\bibnamefont {Amari}}, \bibinfo {author} {\bibfnamefont
  {C.}~\bibnamefont {Medous}}, \bibinfo {author} {\bibfnamefont
  {J.}~\bibnamefont {Paris}}, \bibinfo {author} {\bibfnamefont
  {R.}~\bibnamefont {Hostein}}, \bibinfo {author} {\bibfnamefont
  {J.}~\bibnamefont {Lesueur}}, \bibinfo {author} {\bibfnamefont
  {C.}~\bibnamefont {Feuillet-Palma}},\ and\ \bibinfo {author} {\bibfnamefont
  {N.}~\bibnamefont {Bergeal}},\ }\bibfield  {title} {\bibinfo {title} {{Hybrid
  quantum systems with high-T$_c$ superconducting resonators}},\ }\href
  {https://doi.org/10.1038/s41598-023-41472-z} {\bibfield  {journal} {\bibinfo
  {journal} {Scientific Reports}\ }\textbf {\bibinfo {volume} {13}},\ \bibinfo
  {pages} {14366} (\bibinfo {year} {2023})}\BibitemShut {NoStop}%
\bibitem [{\citenamefont {Mladenov}\ \emph {et~al.}(2022)\citenamefont
  {Mladenov}, \citenamefont {Pankratova}, \citenamefont {Sholokhov},
  \citenamefont {Manucharyan}, \citenamefont {Gross}, \citenamefont {Bushev},\
  and\ \citenamefont {Kukharchyk}}]{Mladenov2022}%
  \BibitemOpen
  \bibfield  {author} {\bibinfo {author} {\bibfnamefont {A.}~\bibnamefont
  {Mladenov}}, \bibinfo {author} {\bibfnamefont {N.}~\bibnamefont
  {Pankratova}}, \bibinfo {author} {\bibfnamefont {D.}~\bibnamefont
  {Sholokhov}}, \bibinfo {author} {\bibfnamefont {V.}~\bibnamefont
  {Manucharyan}}, \bibinfo {author} {\bibfnamefont {R.}~\bibnamefont {Gross}},
  \bibinfo {author} {\bibfnamefont {P.~A.}\ \bibnamefont {Bushev}},\ and\
  \bibinfo {author} {\bibfnamefont {N.}~\bibnamefont {Kukharchyk}},\ }\href
  {https://arxiv.org/abs/2206.03135} {\bibinfo {title} {Microwave cavity-free
  hole burning spectroscopy of
  $\mathrm{Er}^{3+}$\!:$\mathrm{Y}_{2}\mathrm{SiO}_{5}$ at millikelvin
  temperatures}} (\bibinfo {year} {2022}),\ \Eprint
  {https://arxiv.org/abs/2206.03135} {arXiv:2206.03135 [quant-ph]} \BibitemShut
  {NoStop}%
\end{thebibliography}%
	
\end{document}